\documentclass[10pt,letterpaper]{article}
\usepackage[top=0.85in]{geometry}
\usepackage{amsmath,amssymb,relsize,physics}
\usepackage[table]{xcolor}
\usepackage{array}
\usepackage{lastpage,fancyhdr,graphicx}
\usepackage[numbers]{natbib}
\usepackage{hyperref}
\usepackage{authblk}

\usepackage[aboveskip=1pt,labelfont=bf,labelsep=period,justification=raggedright,singlelinecheck=off]{caption}
\newcommand{\olsi}[1]{\,\overline{\!{#1}}}
\newcommand\CoAuthorMark{\footnotemark[\arabic{footnote}]} 

\title{Uncovering potential effects of spontaneous waves on synaptic development: the visual system as a model}

\author[1]{Jennifer Crodelle \footnote{Contributed equally.}}

\author[2,3]{Wei P. Dai \protect\CoAuthorMark \thanks{Corresponding author: weidai@fudan.edu.cn} }

\affil[1]{Department of Mathematics and Statistics, Middlebury College, Middlebury, Vermont, United States of America}
\affil[2]{Research Institute of Intelligent Complex Systems, Fudan University, Shanghai, Shanghai, China}
\affil[3]{Shanghai Artificial Intelligence Laboratory, Shanghai, Shanghai, China}

\begin{document}
\maketitle

\section*{Abstract}
Spontaneous waves are ubiquitous during early brain development and are hypothesized to drive the development of receptive fields (RFs). Different stages of spontaneous waves in the retina have been observed to coincide with the development of the retinotopic map, ON-OFF segregation, and orientation selectivity in the early visual pathway of mammals, and can be characterized by different activity patterns in the retina and downstream areas. Stage II waves, which occur in rodents right after birth, have been implicated in a possible synaptic pruning process, relating these stage II retinal waves to the refinement of the retinotopic map. However, the mechanisms underlying the activity-dependent effects of retinal waves on synapses into the primary visual cortex (V1) are poorly understood. In this work, we build a biologically-constrained model of the development of thalamocortical synapses of V1 cells driven by stage II retinal waves using a spike-timing dependent triplet learning rule. Using this model, together with a reduced rate-based model, we propose mechanisms underlying such a pruning process and predict how characteristics of the retinal waves may lead to different RF structures, including periodic RFs. We introduce gap junctions into the V1 network and show that such a coupling can serve to promote precise local retinotopy. Finally, we discuss how the spatial distribution of synaptic weights at the end of stage II may affect the emergence of orientation selectivity of V1 neurons during stage III waves. The mechanisms uncovered in this work may be useful in understanding synaptic structures that emerge across cortical regions during development.

\section*{Introduction}
Patterned spontaneous activity during development is ubiquitous in the brain \cite{khazipov_early_2006, conhaim_bimodal_2010, ackman_retinal_2012, martini_spontaneous_2021}. The retinal wave is one of such activity that propagates throughout the entire visual system before eye opening\cite{ackman_retinal_2012}. 
Its functional role in setting up synaptic structures along the visual pathway has been the subject of many studies \cite{mclaughlin_retinotopic_2003,grubb_abnormal_2003,pfeiffenberger_ephrin-as_2006,butts_burst-based_2007,ackman_retinal_2012}, yet the exact mechanisms remain unclear. 
In the developing visual system of mice and ferrets, spontaneous waves in the retina occur from before birth (stage I) through the first postnatal week (stage II) to right before eye opening (stage III) \cite{huberman_mechanisms_2008}. These stages of waves are characterized by different mechanisms driving the activity and, as such, display different activity patterns. Stage I waves are mainly propagated through gap-junction connections between retinal ganglion cells (RGCs) \cite{syed_spontaneous_2004,bansal_mice_2000}. Stage II waves, which begin around birth and last through the first postnatal week \cite{huberman_mechanisms_2008}, are dependent upon acetylcholine release from starburst amacrine cells and are characterized by relatively slow wave speeds and restricted domain sizes \cite{feller_requirement_1996}. Stage II waves have been shown to be essential for instructing healthy receptive field refinement \cite{grubb_abnormal_2003,cang_development_2005,stafford_spatial_2009}. Stage III waves, which begin in the second postnatal week and last until eye opening, depend upon glutamate release of retinal bipolar cells and are characterized by highly correlated firing among RGCs of the same sign (ON-ON or OFF-OFF), as well as distinct firing patterns across different RGC types (e.g. alpha, beta, gamma), potentially contributing to distinct wiring patterns across different RGCs \cite{liets_spontaneous_2003, huberman_mechanisms_2008}. In addition, due to the asynchronous and temporally-delayed firing of ON and OFF cells during stage III waves, it is hypothesized that orientation selectivity of V1 cells is mostly set up during this stage of development \cite{thompson_activity-dependent_2017}.   

The correlation between spontaneous retinal waves and circuit development has been studied for years, with the most recent evidence supporting activity-dependent plasticity as underlying circuit refinement during this time \cite{matsumoto_hebbian_2024}. Abolishing or disrupting the retinal waves has been shown to disrupt circuit formation, though the interpretation of these results is often unclear due to other confounding factors \cite{huberman_mechanisms_2008}. Nonetheless, experiments have shown a clear causal link between nicotinic acetylcholine receptors, retinal waves, and the refinements of V1 circuitry \cite{cang_development_2005}. Blocking stage II retinal waves results in large axonal projections and too-large RFs, as does NMDA-blockers in downstream neurons which disrupt the correlated firing needed for synaptic refinement \cite{mclaughlin_retinotopic_2003, grubb_abnormal_2003}. Additionally, in mice whose retinal waves have been blocked genetically, the RGC axonal arbors terminating in the dorsal lateral geniculate nucleus (dLGN) (and superior colliculus) have longer branch lengths that cover a larger area when compared to the wild-type mouse \cite{bansal_mice_2000,pfeiffenberger_ephrin-as_2006,dhande_development_2011}. 

While experiments provide evidence that stage II waves play a role in synaptic refinement, the exact mechanisms underlying the reduction of RF size and the improved acuity of a single neuron remain elusive \cite{thompson_activity-dependent_2017}. In particular, the relationship between characteristics of the retinal waves and the extent of synaptic pruning is unknown. In this work, we build a physiologically plausible computational model of the feed-forward processing in developmental V1 driven by retinal waves to predict how the characteristics of the retinal waves may affect the learned structure of V1 RFs. Together with a rate-based reduced model analysis, we provide the underlying mechanisms to understand these pruning effects.

Beyond the pruning process, this understanding leads to an immediate demonstration on how simple periodic RF structures can arise under certain conditions in our model, which may aid in the understanding of the formation of more sophisticated periodic RFs or connectivity in different brain regions \cite{moser_place_2008,lowel_intrinsic_horizontal_connections_1992} through spontaneous waves \cite{sheroziya_spontaneous_2009,martini_spontaneous_2021}. As one application, we also extend our model to include gap junctions between V1 neurons during the stage II period \cite{li_clonally_2012, yu_specific_2009, molnar_transient_2020} and show that electrical coupling during stage II waves may lead to the refinement of local retinotopy. Other than the retinotopic map, orientation selectivity is one of the most important common feature in V1 across species \cite{ringach_OSmacaque_2002,chapman_development_ferret_1993,niell_highly_2008} that emerges before eye-opening\cite{huberman_mechanisms_2008}. However, little is known about how orientation preference is seeded without experience (likely during stage III \cite{kerschensteiner_glutamatergic_2016, thompson_activity-dependent_2017}), not to mention which developmental characteristics during stage II may affect its emergence.
To shed light on this topic, we apply our model, connecting stage II and III, to first demonstrate that orientation selectivity can indeed be induced by randomly-oriented stage III waves as suggested; and secondly, that its emergence under stage III retinal waves is critically affected by the pruning process during stage II waves. Through these applications, we highlight how the characteristics of spontaneous waves may influence different formations of synaptic structures. 

\section*{Results}
We construct a biologically-motivated model describing spontaneous stage II waves of retinal origin traveling through a local sheet of LGN cells, which in turn project onto cells in V1 with plastic synapses. The waves are considered as randomly-oriented parallel wave fronts that originate distally from the local LGN sheet, motivating our approach to model them as oriented bars. LGN cells are organized on a 16 $\times$ 16 grid (corresponding to a $20\times20$ degree visual field in mouse), with ON and OFF cells co-located at each vertex \cite{reid_LGN-V1_1995} (which respond to light and dark stimuli, respectively), and output spikes in response to spatiotemporal activation from the waves through a linear-nonlinear-Poisson process. Individual V1 cells receive synaptic input from 95\% of randomly-selected LGN cells and are modeled as adaptive exponential integrate-and-fire neurons\cite{brette_adaptive_2005}. See Figure \ref{fig:RF_refinement}A for a model schematic and \nameref{sec:models_and_methods} for more details on the model setup. 

The weights of the synaptic connections from LGN to V1 are plastic, following the triplet spike-timing dependent plasticity rule with parameters fit to visual cortex\cite{pfister_triplets_2006,zenke_synaptic_2013}. To start, V1 cells are uncoupled \cite{yu_specific_2009}, essentially treated as random realizations or trials in the model. Gap junctions among V1 neurons are considered later in the paper. See \nameref{sec:models_and_methods} for more details on the model equations and parameters.

To study the receptive field (RF) development of V1 neurons during stage II waves, we present retinal waves to the local sheet of LGN cells and measure the RF of each V1 cell as the collection of potentiated synapses from LGN to V1. Note that we use RF, feedforward RF and feedforward LGN connectivity interchangeably in this work.

\subsection*{Stage II waves guide the synaptic pruning of LGN to V1 connections}

\begin{figure}[ht!]
    \centering
    \includegraphics[scale=0.7]{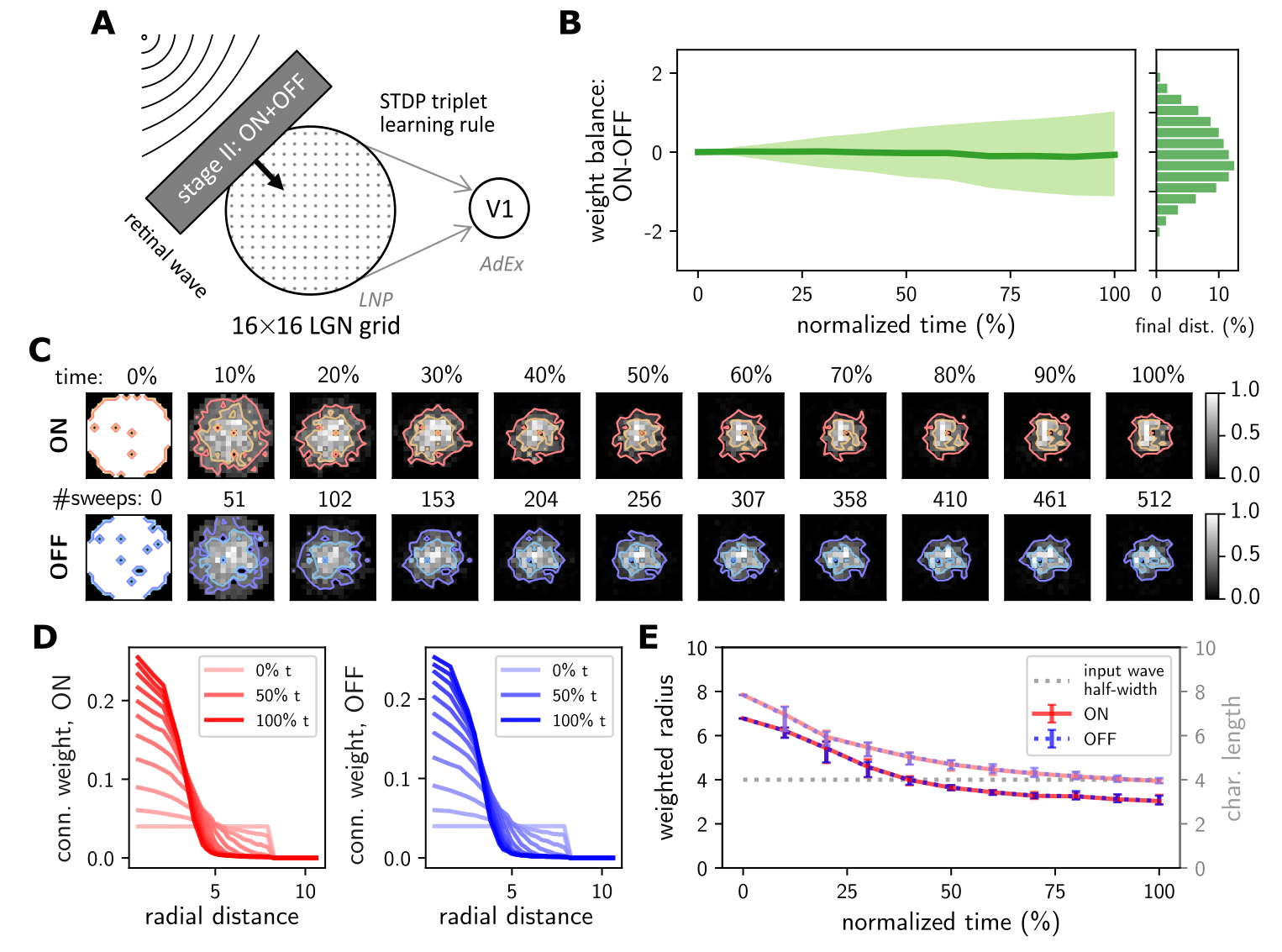}
    \medskip
    \caption{{\bf Model setup and typical receptive field refinement}. {\bf A.} Schematic of the stage II wave as a random parallel wave front that activates the LGN cells. Synapses from LGN to all V1 cells are plastic, following a triplet STDP learning rule. {\bf B.} Temporal structure (left) and final distribution (right) of the difference between the connection strength from ON and OFF LGN cells, averaged across all 1024 V1 neurons. Shading represents standard deviation. {\bf C.} The ON (top) and OFF (bottom) RF refinement of one sample V1 neuron over time shown by the spatiotemporal change of connection weights in grayscale (normalized by local maximum). The red (ON) and blue (OFF) solid-line contours enclose the potentiated connections while the orange (ON) and navy (OFF) contours trace the local half-maximum. {\bf D} Average ON (left, red) and OFF (right, blue) connection weight distribution along the radial axis (averaged over polar angles) over time (saturating color) averaged across 24 trials. {\bf E.} (left y-axis, saturated colors) Temporal evolution of radius-weighted ON (solid red) and OFF (dotted blue) LGN connection weights averaged across all 1024 V1 neurons and (right y-axis, less saturated colors) the characteristic length averaged across all 1024 V1 neurons. The error bars show the standard deviation in the average over 24 trials. The gray dotted line is at the half-width of the input wave.}
    \label{fig:RF_refinement}
\end{figure}

We begin by demonstrating that, with parameters for wave speed and width chosen near the average for the range of experimentally-determined values \cite{maccione_following_2014}, our model simulations result in a refinement of the RF of V1 cells. The simulation begins with 1024 uncoupled V1 neurons receiving plastic excitatory input from a circular patch of the LGN sheet, mimicking the reachable pool of LGN cells through their local spread of axonal terminals; recall Figure \ref{fig:RF_refinement}A. In response to many presentations of stage II retinal waves, we record how the synaptic weights from LGN cells change over time and measure the RF as the radius of the collection of potentiated synapses (see \nameref{sec:models_and_methods} for more details on RF measures). 

As expected, due to the simultaneous activation of ON and OFF LGN cells during stage II \cite{kerschensteine_precisely_2008}, the distribution of the connection weight difference between ON and OFF LGN cells to one V1 neuron is balanced on average over time; see Figure \ref{fig:RF_refinement}B. Synapses from LGN cells near the center of the local sheet potentiate over time while those in the surrounding regions decay, resulting in a disc-shaped RF structure for both ON and OFF LGN cells; see Figure \ref{fig:RF_refinement}C. Recording the weight of the ON (Figure \ref{fig:RF_refinement}D) and OFF (Figure \ref{fig:RF_refinement}E) synapses into one V1 cell across the radial axis and time (averaged over 24 trials) shows that synaptic weights at small radial distances, near the middle of the LGN pool, potentiate while those in the periphery decay. Averaging over time, we quantify this shrinking of the disc-like RF using two measures: 1) connection-weighted radius: the average distance to RF center weighted by connection strength and 2) characteristic length: the half-width of the contoured area as measured by the square root of the area of potentiated weights divided by two (see \nameref{sec:models_and_methods} for more details); see Figure \ref{fig:RF_refinement}F. We note here that synapses from ON and OFF LGN cells behave identically, consistent with experiments showing that ON-OFF segregation does not occur until near the end of stage II retinal waves \cite{huberman_mechanisms_2008}. As such, we refer to the RF of both ON and OFF cells as just ``RFs" unless otherwise specified.

\subsection*{The effect of wave speed and width on V1 RF}
Stage II waves in experiments show a large variation in speed and width \cite{maccione_following_2014}. In the model, we vary the speed and width of incoming retinal waves and characterize the effects on the RF structure.

\subsubsection*{Wave speed}
The wave speed, $v$, is varied from $2$ deg/$s$ (half the default speed) to $16$ deg/s (four times the default speed) while all other parameters remain the same.

\begin{figure}[ht!]
    \centering
    \includegraphics[width=3.9in]{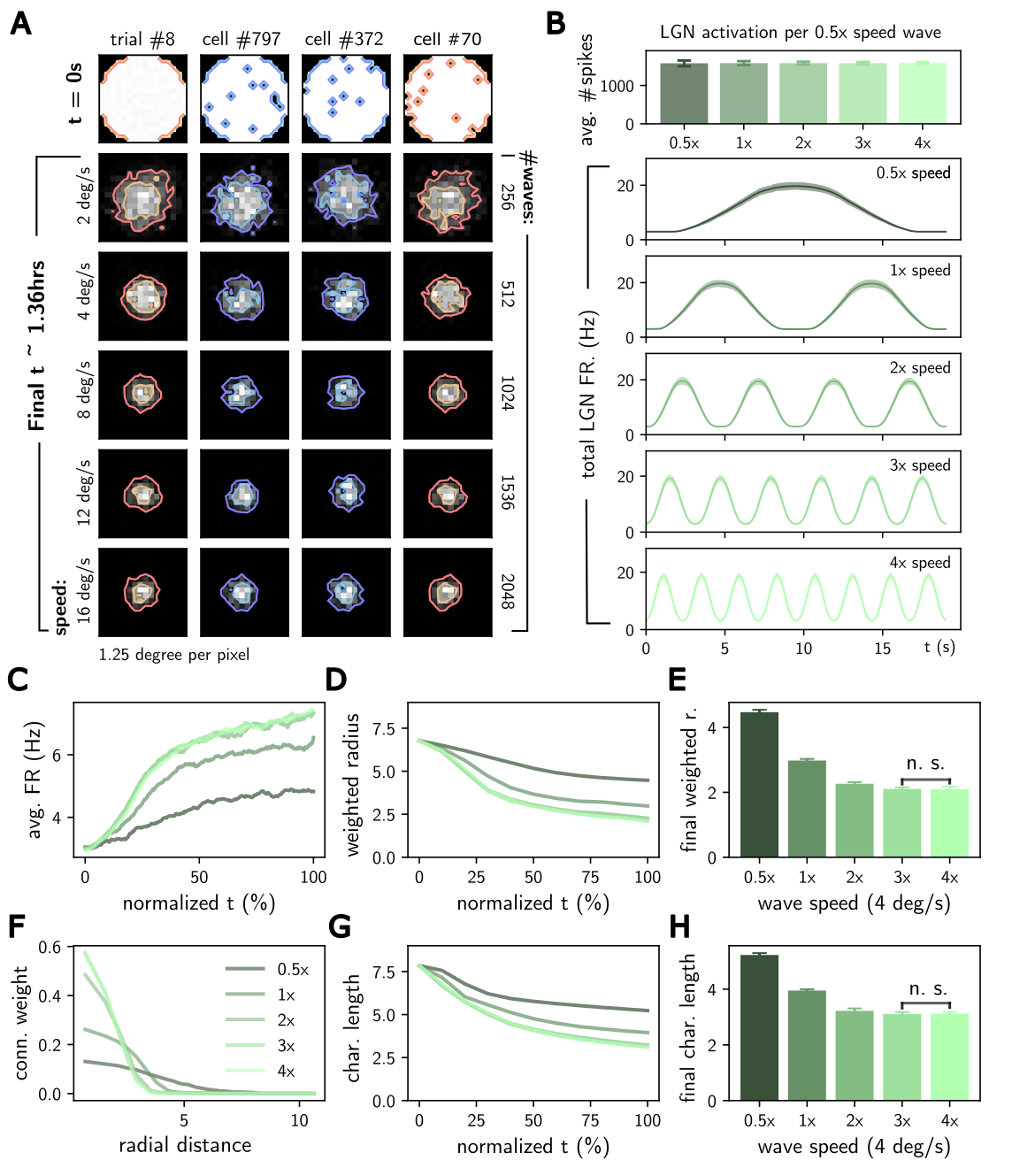}
    \medskip
    \caption{{\bf Effect of wave speed.} {\bf A.} The LGN-V1 connection profiles (ON in red, OFF in blue) with different wave speeds averaged across all 1024 neurons in one trial (left-most column) and three individual neurons within one trial (three right-most columns). The top row shows available connections at $t=0$, while the remaining rows show the final weight distribution for increasing wave speeds. {\bf B.} The LGN activation, measured as the number of LGN spikes per normalized time range, remains the same (T-test p-value = 0.415) across wave speeds (speed is colored with increasingly saturated green). Lower panels show the firing rate of all connected LGN cells, averaged over retinal waves. {\bf C.} Averaged V1 firing rates over time (81.7 minutes in total) across wave speeds. {\bf D.} Averaged connection-weighted radius of LGN connections over time across wave speeds. {\bf E.} Weighted radius at the end of the simulation across wave speeds. T-test cannot distinguish between $3\times$ and $4\times$ the typical wave speed (4 deg/s), p-value = 0.4966. {\bf F.} Radial distribution of final connection weights (averaged over discrete polar angles) across wave speeds. {\bf G.} Characteristic length over time across wave speeds. {\bf H.} Characteristic length at the end of the simulation across wave speeds. T-test cannot distinguish between $3\times$ and $4\times$ the typical wave speed (4 deg/s), p-value = 0.4253. Error bars in E and H denote the 10\%-90\% range of values from 18 trials. Green lines and bars indicate averages across both ON and OFF subregions (or LGN cells) across all V1 neurons.}
    \label{fig:wave_speed}
\end{figure}

Results show that the size of the disc-like RF shrinks as wave speed increases. This occurs consistently across all V1 cells in one sampled trial (Figure \ref{fig:wave_speed}A, first column), as well as for both ON and OFF LGN input to sampled cells (Figure \ref{fig:wave_speed}A, final 3 columns). While the number of waves in a fixed time range increases proportionally with wave speed, the total LGN activation, measured by the average number of LGN spikes in the period of the slowest wave remains relatively constant across wave speeds; see Figure \ref{fig:wave_speed}B. Nonetheless, the overall firing rate of the V1 cells increases with wave speed while the rising and saturating phases of the V1 firing rate, indicating periods of learning, are matched across speeds; see Figure \ref{fig:wave_speed}C. 

Over time, the connection-weighted radius of the disk-like RF decreases over time for all wave speeds, illustrating a pruning process, with the greatest decrease occurring for the two fastest wave speeds; see Figures \ref{fig:wave_speed}D and E. Not only does increasing wave speed result in a smaller disk-like RF, but the connection weights at the center of the RF are larger for fast wave speeds, decreasing dramatically as the radial distance increases; see Figure \ref{fig:wave_speed}F. These results are supported by measurements of the characteristic length, which shows a significant decrease as wave speed increases, although notably there is no significant difference between RFs resulting from the two fastest speeds; see Figures \ref{fig:wave_speed}G and H. 


\subsubsection*{Wave width}
In this subsection we vary the width, $w$, of the Stage II wave from $4$ LGN cells (half the default width) to $12$ LGN cells (1.5 times the default width). The waves are presented at random orientations at the default wave speed ($4$ deg/s) and we measure the resulting RF structure. 

\begin{figure}[ht!]
    \centering
    \includegraphics[width=4.7in]{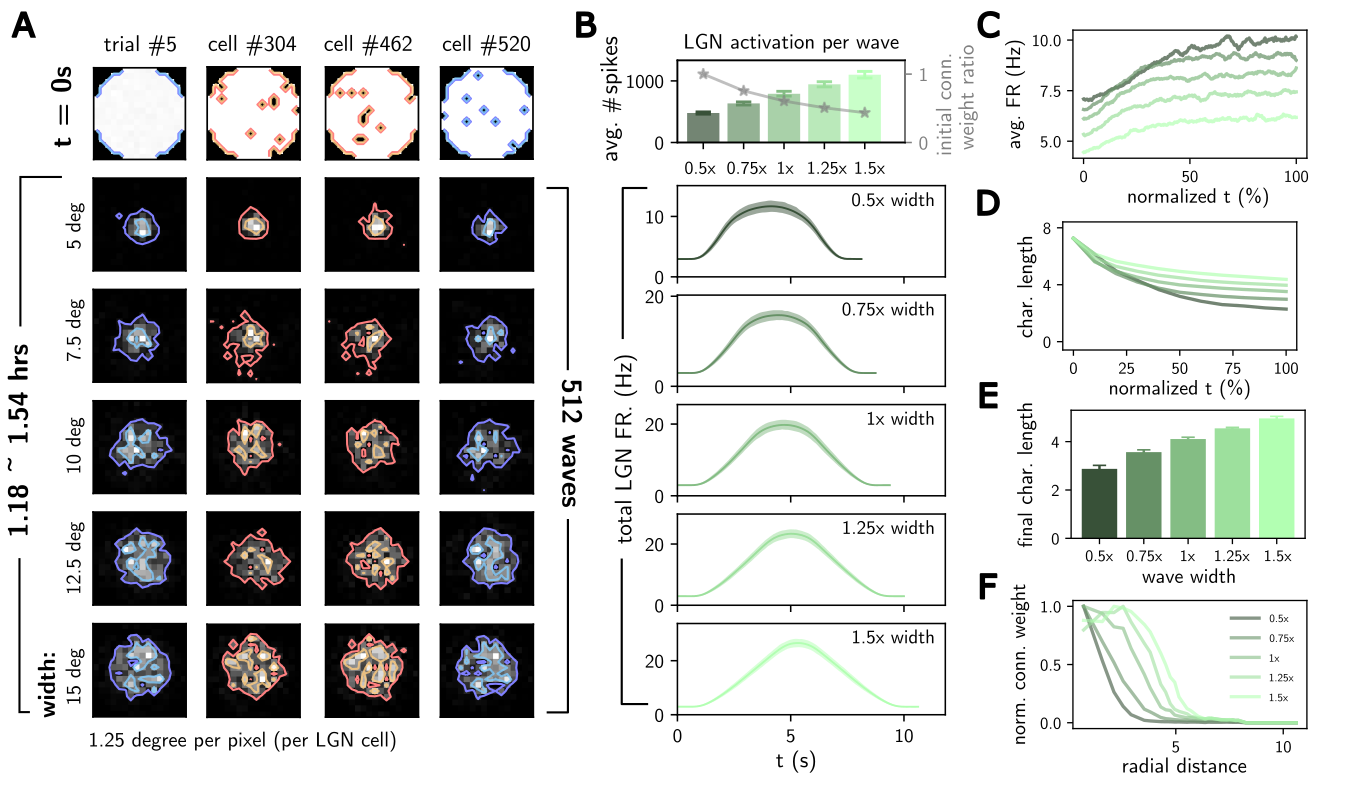}
    \caption{{\bf Effect of wave width.} {\bf A.} Averaged (left-most column) and samples (three right-most columns) of 2-dimensional LGN-V1 connection profiles (ON in red, OFF in blue) with varying wave widths. The top row shows available connections at $t=0$, while the remaining rows show the final weight distribution for increasing wave width. {\bf B.} Top panel shows the total LGN spikes per wave across wave widths (green bars, color coded with increasingly saturated green) and initial LGN-V1 connection strength, normalized to the smallest wave width (gray stars). Lower panels show the wave-averaged firing rate of all connected LGN cells across wave widths. {\bf C.} Averaged V1 firing rates over time. {\bf D.} Characteristic length over time across wave widths. {\bf E.} Characteristic length at the end of the simulation across wave widths. {\bf F.} Normalized radial distribution of final connection weights (averaged over discrete polar angles) across wave widths. Error bars denote the 10\%-90\% range of values from 18 trials.}
    \label{fig:wave_width}
\end{figure}

As wave width increases, the size of the disk-like RF increases, as measured across all V1 cells in an individual trial (Figure \ref{fig:wave_width}A, first column), as well as for both ON and OFF LGN input to individual cells (Figure \ref{fig:wave_width}A, final 3 columns). 
In this case, larger wave widths activates more LGN cells per unit time; see Figure \ref{fig:wave_width}B. To adjust for this change across wave width, we decrease the initial connection strength of the LGN synapses directly in proportion to the average number of spikes induced by different wave widths to keep LGN activation the same (see gray stars in Figure \ref{fig:wave_width}B and \ref{SIfig:width_same_strength} for results with constant initial connection strength). The resulting rising and saturating phases of the firing rate of the V1 cell are matched across different widths; see Figure \ref{fig:wave_width}C.

The characteristic length of the potentiated area decreases over time for all widths, again illustrating a pruning process. The greatest decrease occurs for the smallest wave width; see Figures \ref{fig:wave_width}D and E. The connection weights for the smallest width show the most dramatic decrease with radial distance; see Figure \ref{fig:wave_width}F. 

To achieve a in-depth understanding of the mechanisms underlying the observed refinement of RFs, we introduce a reduced firing rate-based version of this model in the next section and use it to identify the main mechanisms contributing to RF refinement across wave speed, width and more. Leveraging the mechanistic understanding of RF refinement in the reduced model across several applications, we illustrate the following results: i) the emergence of a ring-like, or even periodic RF structure is dependent upon initial connectivity strength and the amount of long-term depression in the plasticity model (see Figure \ref{fig:w0_LTD}); ii) the presence of gap junctions with RF refinement during stage II plays an important role in maintaining local retinotopy along with theRF refinment(see Figure \ref{fig:retinotopy_refinement}); and iii) the size and shape of the RF after stage II waves determines the orientation selectivity that arises during stage III waves (see Figure \ref{fig:stage3OS}).

\subsection*{A reduced rate-based model of V1 RF development}
\label{sec:reduced_model}
\begin{figure}[ht!]
 \centering
\includegraphics[width=5.1in]{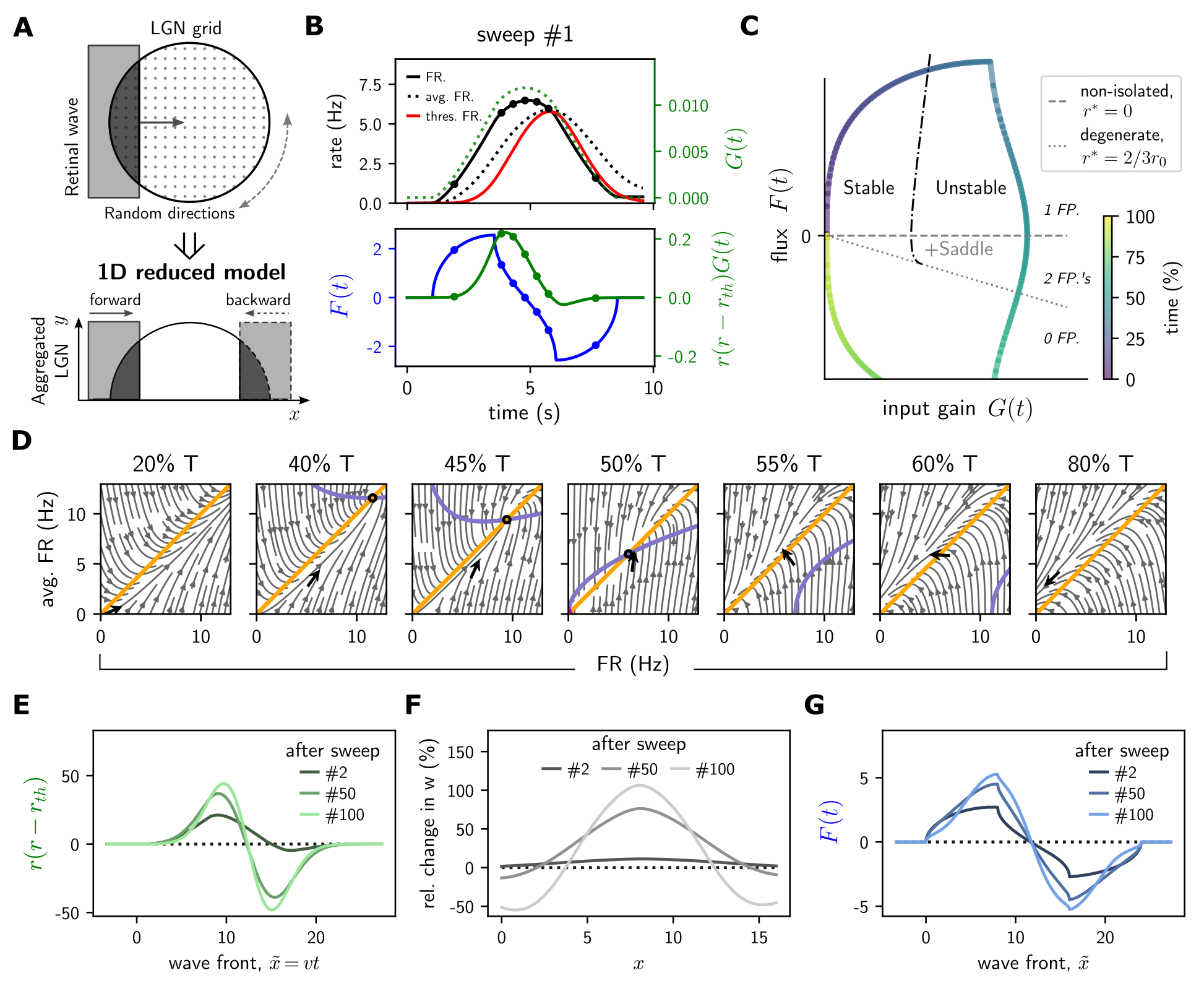}
    \caption{ {\bf Rate-based reduced model.} 
    {\bf A.} Schematic illustrating the 1-D reduced model derived from the full simulation. 
    {\bf B.} Temporal evolution of the system during the first wave.
        The upper panel shows the V1 firing rate (solid) and its average (dotted) in black, the threshold firing rate, $r_{\text{th}}$, in red.
        The input gain, $G(t)$, in dotted green is plotted against the right y-axis.
        The lower panel shows the two terms on the right-hand-side of $dr/dt$ in Eq. (\ref{eqn:reduced1}).
    {\bf C.} Phase diagram based on changes in $G(t)$ and $F(t)$ with parameters chosen to illustrate all possible regimes.
        Regimes of single, double and zero fixed points (FP.'s) are separated by a dashed gray line of non-isolated FP.'s, and a dotted gray line of degenerate FP.'s, in-between which lies the saddle points. 
        The stable and unstable regimes are separated by the dash-dotted line.
    {\bf D.} Phase portraits of $r(t)$ vs $\olsi{r}(t)$ at seven time points during the first wave corresponding to solid circles in B.
        The $\olsi{r}$-nullclines are in orange and the $r$-nullclines are in purple.
        The black arrows mark the vectors ($r$, $\olsi{r}$ at the corresponding time points.
    {\bf E.} Temporal change in $r(r-r_{\text{th}})$ after 2, 50, and 100 wave sweeps plotted against the wave front, $\tilde{x} = vt$.
    {\bf F.} Change in the LGN connection weights relative to their initial values in percentage across LGN space, $x$.
    {\bf G.} Flux term $F(t)$ plotted against the wave front. 
    Note, to exemplify the effect of the plasticity rule alone, global homeostasis is turned off for these panels.}     
    \label{fig:reducedModel_intro}
\end{figure}

Here we introduce a reduced firing-rate model to aid in understanding potential mechanisms underlying the full model-predicted changes in RF structure. In this model, we collapse the LGN onto a one-dimensional space parameterized by $x$, along which retinal waves of width $2d$ may be presented from left to right or right to left; see Figure \ref{fig:reducedModel_intro}A. The reduced model comprises three equations: 1) the change in firing rate of the V1 cell, $r(t)$; 2) the change in average firing rate of the V1 cell, $\bar{r}(t)$; and 3) the change in connection weight from a position, $x$, in the one-dimensional LGN space to the V1 neuron, $w(x,t)$. Following previous approaches to deriving a BCM-like mean-field approximation to the triplet learning rule \cite{zenke_synaptic_2013} and applying reasonable assumptions on the relative rate of change of the weight compared to the firing rate (see \nameref{appendix:A} for full reduced model derivation), we consider the resulting two-dimensional ODE model for the firing rate and average firing rate below:

\begin{eqnarray}
    \label{eqn:reduced1}
    \left\{\:
    \begin{aligned}
    \frac{\dd r}{\dd t} & = r\left(r-r_{\text{th}}\right) G\left(t\right) + F\left(t\right)\\
    \tau\frac{\dd \olsi{r}}{\dd t} &= -\olsi{r} + r
    \end{aligned}\right.\kern-\nulldelimiterspace,
\end{eqnarray}
where $r_{\text{th}}$ is the threshold firing rate that determines the sign of the input contribution to firing rate.

The equation for the change in average firing rate, $\bar{r}(t)$, is straightforward.
The contribution to the dynamics of the firing rate, $r(t)$, consists of two parts: the weight change due to learning within an activated area of LGN, described by the input gain $G(t)$, and the weight change on the activation boundaries of the wave, described by the flux $F(t)$. 
Integrating the LGN input over the activated part of the pool, we can write down the equations for the input gain and flux as follows 
    \begin{eqnarray}
        \label{eqn:r_F_G}
        \left\{\:
        \begin{aligned}
        G(t) & = g_0\:A^{\scriptscriptstyle +}\:\tau_{\scriptscriptstyle +}\tau_{\scriptscriptstyle slow}\hat{r}^{2}\int_{b(t)}^{u(t)}g(x)\dd x \\
	F(t) & = g_0\:\hat{r}\: v \left[ g(x)w(x)\right]_{\partial \Omega\left(t\right)}, 
        \end{aligned}\right.\kern-\nulldelimiterspace
    \end{eqnarray}
where $g_0$ is the intrinsic gain of the neuron, $v$ is the speed of a wave with half-width $d$ traveling in a pool of LGN cells with radius $l$ and firing rate $\hat{r}$, when activated. The flux depends on evaluating the LGN activation curve, $g(x)$, and the synaptic weights, $w(x)$ at the boundary of the wave, $\partial \Omega$ defined by $u(t)$ and $b(t)$ (see details in \nameref{appendix:A}). Assuming synaptic weight changes are on a slower time scale than firing rate changes, we write the simplest version of the reduced model which updates the weights after the $i$-th presentation of a wave, $w_i(x)$, as follows
\begin{equation}
        w_{i}(x) = w_{i-1}(x) + \tau_{\scriptscriptstyle +}\tau_{\scriptscriptstyle slow}A^{\scriptscriptstyle +}\hat{r}\int_{x/v}^{(x+2d)/v}r\left(r-r_{\text{th}}\right)dt.
             \label{eqn:approx_weight_update}
\end{equation}
The parameters $\tau_{\scriptscriptstyle +}$, and $\tau_{\scriptscriptstyle\text{slow}}$ come from the STDP triplet rule (see \nameref{sec:models_and_methods}).
Here, the rate-based weight dynamics is BCM-like as it involves a threshold firing rate, $r_{\text{th}}=\left(\olsi{r}/r_0\right)\olsi{r}$, which upon crossing changes the sign of weight change \cite{gjorgjieva_triplet_2011, zenke_synaptic_2013} ($r_0$ is the target firing rate that determines the amplitude of synaptic depression in the triplet rule).
The assumption to keep weights constant during the wave presentation allows for a more analytically tractable two-dimensional system of ODEs; however, throughout this work, we show that the results for the full system (which includes the partial differential equation for the weight change) qualitatively match those for the simplified two-dimensional model since the learning rate, $A^{\scriptscriptstyle{+}}$, is small, $1\times10^{-3}$.

To illustrate the dynamics of the reduced model, we plot the system variables, $r,\olsi{r},r_{\text{th}}$ including the input gain, $G(t)$, and flux $F(t)$, during the presentation of the first wave (``sweep \#1'', see Figure \ref{fig:reducedModel_intro}B).
Notice that the firing rate of the V1 cell is mainly driven by the change in flux, which is one magnitude stronger than the input gain multiplied by the change in weights (see right y-axis of the bottom panel in Figure \ref{fig:reducedModel_intro}B) due to the small learning rate. 
As the wave sweeps past the center, the flux switches sign and input gain reaches the peak, while the firing rate reaches its own peak slightly later.

The temporal phase diagram of the flux and input gain in Figure \ref{fig:reducedModel_intro}C demonstrates all possible regimes that the system can achieve during the first wave.
The temporal trace of $F(t)$ and $G(t)$ is symmetric since the weight is fixed until the end of the wave by assumption; however, the experienced regimes are asymmetric during the rise and decay of the response (see panels in Figure \ref{fig:reducedModel_intro}D).
The system begins with a single fixed point (FP.), which can be unstable if the input gain is too strong, that raises the firing rate (first three panels in Figure \ref{fig:reducedModel_intro}D). Then, it transitions a regime with two fixed points (at and immediately after 50\%T where a saddle FP is born from a non-isolated FP at $(0,0)$). 
Then, the $r$-nullcline detaches from the $\olsi{r}$-nullcline at a bifurcation point where the two FP's merge into a degenerate FP ($r^{*}=2/3r_0$) and finally enters the regime with zero fixed points, but an absorbing boundary at $r = 0$ (the three panels in Figure \ref{fig:reducedModel_intro}D to the right).

As the number of wave sweeps increase (from 2, 50 to 100 sweeps, alternating forward and backward), the effect size of threshold crossing is amplified, as shown in Figure \ref{fig:reducedModel_intro}E.
More importantly, the synaptic change, proportional to $r\left(r-r_{\text{th}}\right)$, experiences a much more significant change in depression over potentiation, and the crossing happens closer to the center of the LGN pool.
Consequently, the relative change of the synaptic weight is a bell-shape function that sharpens as the number of sweeps increases; see Figure \ref{fig:reducedModel_intro}F.
We further observe that both the relative change in synaptic weight and the flux term quickly rise and then saturate from sweep \#50 to \#100 (Figure \ref{fig:reducedModel_intro}E and G).
Altogether, it is clear that, at first, the asymmetric gaps between $r$ and $r_{\text{th}}$ cause the early potentiation in synaptic weight including the center part  of the LGN pool (Figure \ref{fig:reducedModel_intro}B), resulting in a faster and stronger change in the flux term (notice the change of slope near the sign change in Figure \ref{fig:reducedModel_intro}G).
Then in turn, the flux term, as the main driver to $dr/dt$, induces a faster and stronger threshold crossing of $r_{\text{th}}$ right after the wave sweeps past the center part, pushing the boundary of potentiation and depression closer to the center.
Finally, since the waves alternate in both direction, the sharpening effect happens on both sides, rendering the spatial distribution of synaptic weight a bell-shaped function.

Now that we have a basic understanding of the dynamics of the reduced model under default parameters, we use the model to replicate the results in our full model simulations with changes in the wave speed and wave width to uncover the mechanisms underlying the corresponding RF refining process.

\subsubsection*{Mechanisms underlying the effect of changes in wave speed and width}
\begin{figure}[ht!]
 \centering
    \includegraphics[width=5in]{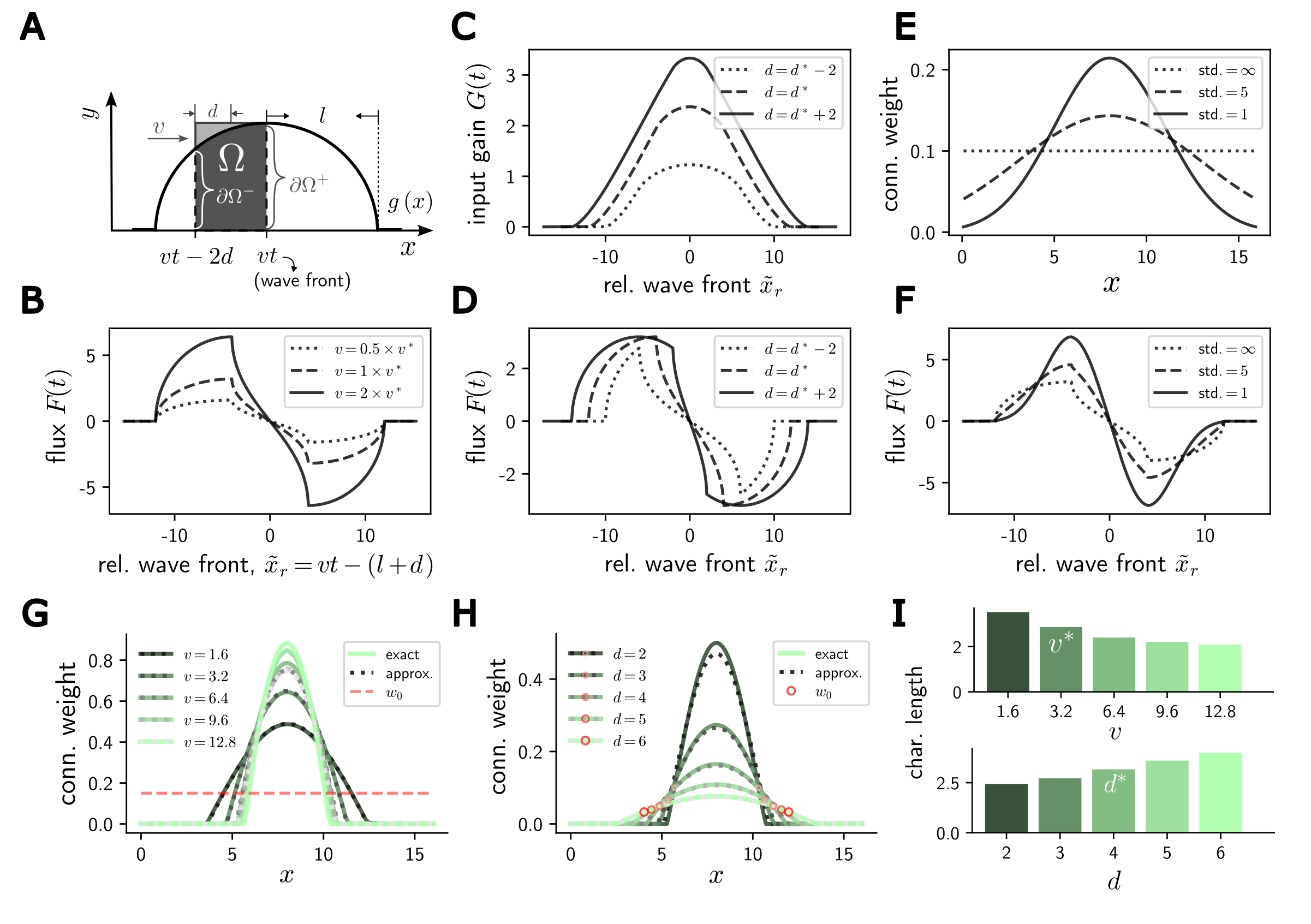}
    \caption{ {\bf Mechanisms underlying effects of wave speed and width}. {\bf A.} Schematic illustrating a wave with speed $v$, half-width $d$, and boundary $\partial\Omega$ traveling through a 1-dimensional continuous LGN space with half-length $l$. {\bf B.} The flux $F(t)$ plotted against relative wave front over time (for alignment at the center) across wave speeds. {\bf C} and {\bf D.} Input gain $G(t)$ and the flux $F(t)$ plotted against relative wave front across wave widths, respectively. {\bf E.} Spatial distribution of LGN connection weights across different numbers of intermediate wave sweeps illustrated using Gaussian functions (see Figure \ref{fig:reducedModel_intro}F for actual numerical result). {\bf F.} The flux $F(t)$ across different weight distributions given in E. {\bf G.} Spatial weight distribution of the reduced model are plotted for $[50, 100, 200, 300, 400]$ sweeps with the increasing wave speeds, $v$, accordingly, making sure the activation durations are keeps the same. The exact solutions are denoted by solid green lines, and the approximated solutions are denoted by dotted gray lines. Increment in the wave speeds are denoted by the increment in lightness. The red dashed line mark the initial connection weight, $w_0 = 0.15$. {\bf H.} Same as G, but for change in the wave width, $d$, and different initial weights, $w_0\simeq[0.08,0.06,0.05, 0.04,0.035]$, are marked in red circles. {\bf I.} The characteristic length at the end of the simulation across wave speed (top) and width (bottom).}
    \label{fig:mechanism}
\end{figure}

In this section, we mimic the simulation experiments for changing wave speed and width in the reduced model to explore the main mechanisms contributing to the observed effects. Figures \ref{fig:mechanism}A-D illustrate the effect of changing wave speed, $v$, and wave width, $d$, on the LGN input gain and activation, $F(t)$ and $G(t)$, respectively, after one wave presentation. The predominant effect of changing wave speed can be seen as a vertical stretching of the flux; see Figure \ref{fig:mechanism}B, which comes from the linear dependence of $F(t)$ on $v$ in Eq.~(\ref{eqn:r_F_G}). Increasing the wave width, $d$, however, increases the LGN input gain while the flux across different wave widths remains of similar amplitude, but has wider peaks and valleys for larger wave widths; see Figures \ref{fig:mechanism}C and D.  

As the height of the peak in connection strength increases in the middle of the LGN with the number of waves presented (Figure \ref{fig:mechanism}E), the flux $F(t)$, shows some stretching and shifting toward the center; see Figure \ref{fig:mechanism}F. Taken together, the effect of increasing wave speed on the learned weight distribution across a one-dimensional LGN space is a significant increase in peak height at the center of the LGN space and a shrinking of peak width, driven primarily by the flux; see Figure \ref{fig:mechanism}G. Conversely, the effect of increasing the wave width on the learned weight distribution is a decrease in the amplitude of the peak at the center of the LGN space and a spreading of the width, driven primarily by input gain, with some effects from the flux; see Figure \ref{fig:mechanism}H. Results from the reduced model are similar to results from the full simulation (compare Figures \ref{fig:wave_speed} and \ref{fig:wave_width} to Figure \ref{fig:mechanism}I), while also providing  intuition behind the main mechanisms leading to the observed trends.

\subsubsection*{Formation of a ring-like receptive field}

\begin{figure}[ht!]
\centering
    \includegraphics[width=5.2in]{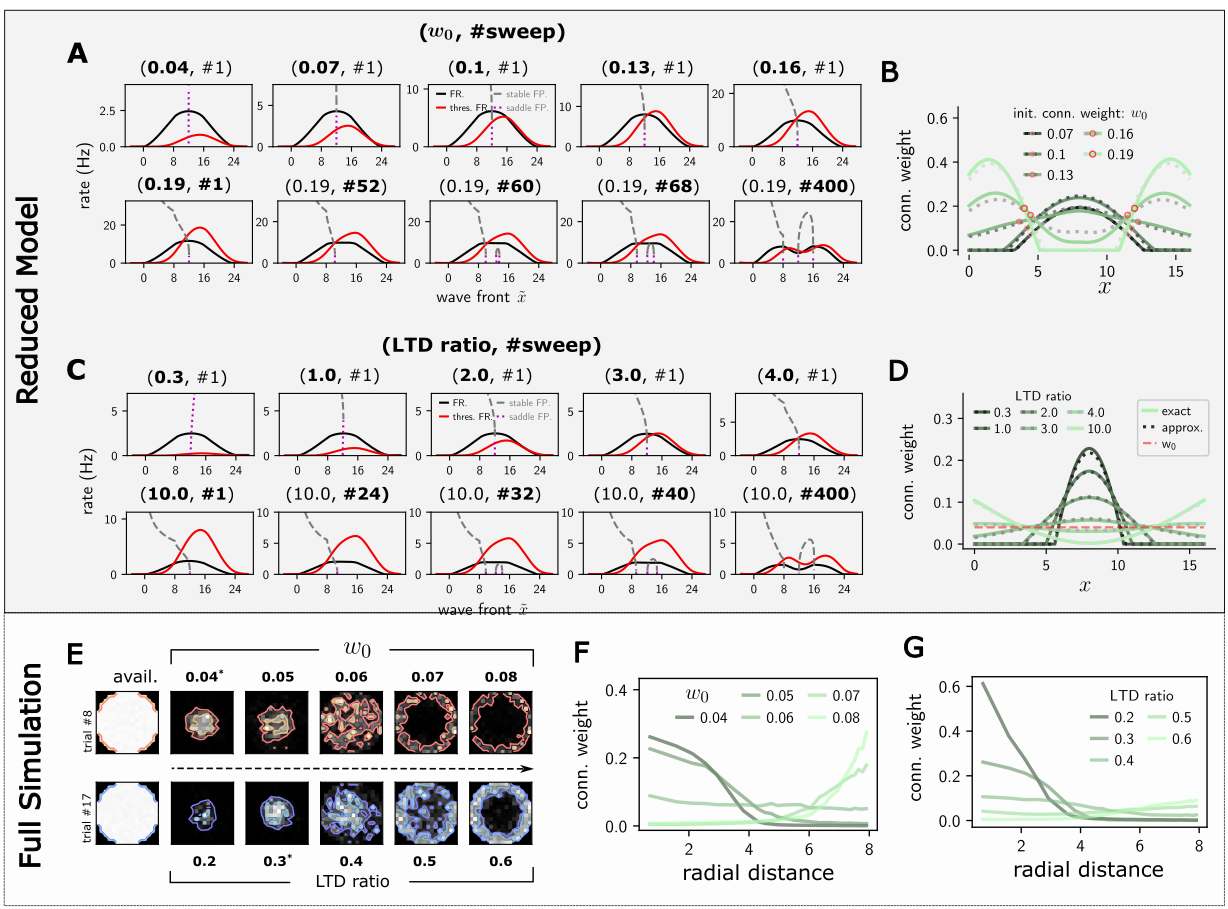}
    \medskip
    \caption{ {\bf Emergence of a ring-like structure.} 
    {\bf A.}  Average V1 firing rate (black curve) and threshold firing rate (red curve) plotted against wave front after the initial wave sweep across increasing initial weights (top row) and for a high initial weight, $w_0 = 0.19$, across an increasing number of wave sweeps (bottom row). Gray dashed line shows the existence of a stable FP while pink dotted line shows the emergence of a saddle point.
    {\bf B.} Results from 400 sweeps of waves in alternating direction from the exact (solid green) and approximated (dotted gray) reduced model. 
    {\bf C.} Same as A, but for increasing LTD ratio (top row) and for a fixed LTD ratio, $r_{\text{LTD}} = 10$, across an increasing number of wave sweeps (bottom row). The radius of the LGN pool is 8 and wave width is 4.
    {\bf D.} Same as B, but for the LTD ratio.
    The initial connection weights are the same for different LTD ratios (dashed red).
    {\bf E.} Two trial-averaged samples from the full simulation as the initial weight increases (top row) and LTD ratio increases (bottom row). The asterisks denote the default parameters. The first column shows the available connections. {\bf F.} Radial distribution of the final connection weights in the corresponding full simulations in E. 
    {\bf G.} The same as F, but for LTD ratio.
    }
    \label{fig:w0_LTD}
\end{figure}

Here we use our understanding of the reduced model to uncover two mechanisms underlying the transition from a disk-like RF to a ring-like RF.
Recall from Figure \ref{fig:reducedModel_intro} that the fixed point goes from stable, driving the firing rate toward a large positive value for $r(t)$, to unstable, driving the firing rate back down toward zero. The location of the stable fixed point, as well as its transition to an unstable saddle point, depends upon the location of the crossing of $r(t)$ and $r_{th}$. As $w_0$ increases, $r(t)$ increases and widens, shifting the intersection point to the left and lowering the height of the stable fixed point during the first sweep; see Figure \ref{fig:w0_LTD}A, top row. At the largest tested $w_0$ value, across numerous presentations of waves, the crossing is shifted enough to the left such that the fixed point goes from stable to unstable (saddle) twice, creating two small peaks in the firing rate curve; see Figure \ref{fig:w0_LTD}A, bottom row. 

Due to the $r - r_{\text{th}}$ term in Eq.~(\ref{eqn:approx_weight_update}), the crossing of $r$ and $r_{\text{th}}$ also determines the difference between long-term potentiation (LTP) and depression (LTD) in the weights. 
Increasing the initial connection weight initially results in a wider disk-like RF; see three darkest green curves in Figure \ref{fig:w0_LTD}B. Eventually, however, for large enough initial weight, LTD is induced in the center of the LGN, leaving peaks in the connection weight on the ends of the LGN; see three lighter curves in Figure \ref{fig:w0_LTD}B.

A similar trend is observed when the ratio of LTD to LTP, termed the LTD ratio, is increased, accomplished by scaling $A^+$ in Eq (\ref{eqn:r_F_G}) and (\ref{eqn:approx_weight_update}); see Figures \ref{fig:w0_LTD}C and D. Notice from Eq (\ref{eqn:approx_weight_update}) that increasing $A^+$ is identical to decreasing $r_0$, the target firing rate, which results in increasing the threshold, $r_{th}$. Increasing $r_{th}$ results in a similar left-ward shift of the intersection of $r(t)$ and $r_{th}$ as observed in the case of changing $w_0$. 

This result is replicated in the 2-dimensional full simulation as well. Increasing $w_0$ and the LTD ratio results in the formation of a ring-like RF, with potentiation of the weights along the outskirts of the RF and depression in the center; see Figure \ref{fig:w0_LTD}E-G. 

\subsection*{Application: Emergence of periodic RFs}
\begin{figure}[ht!]
\centering
    \includegraphics[width=5in]{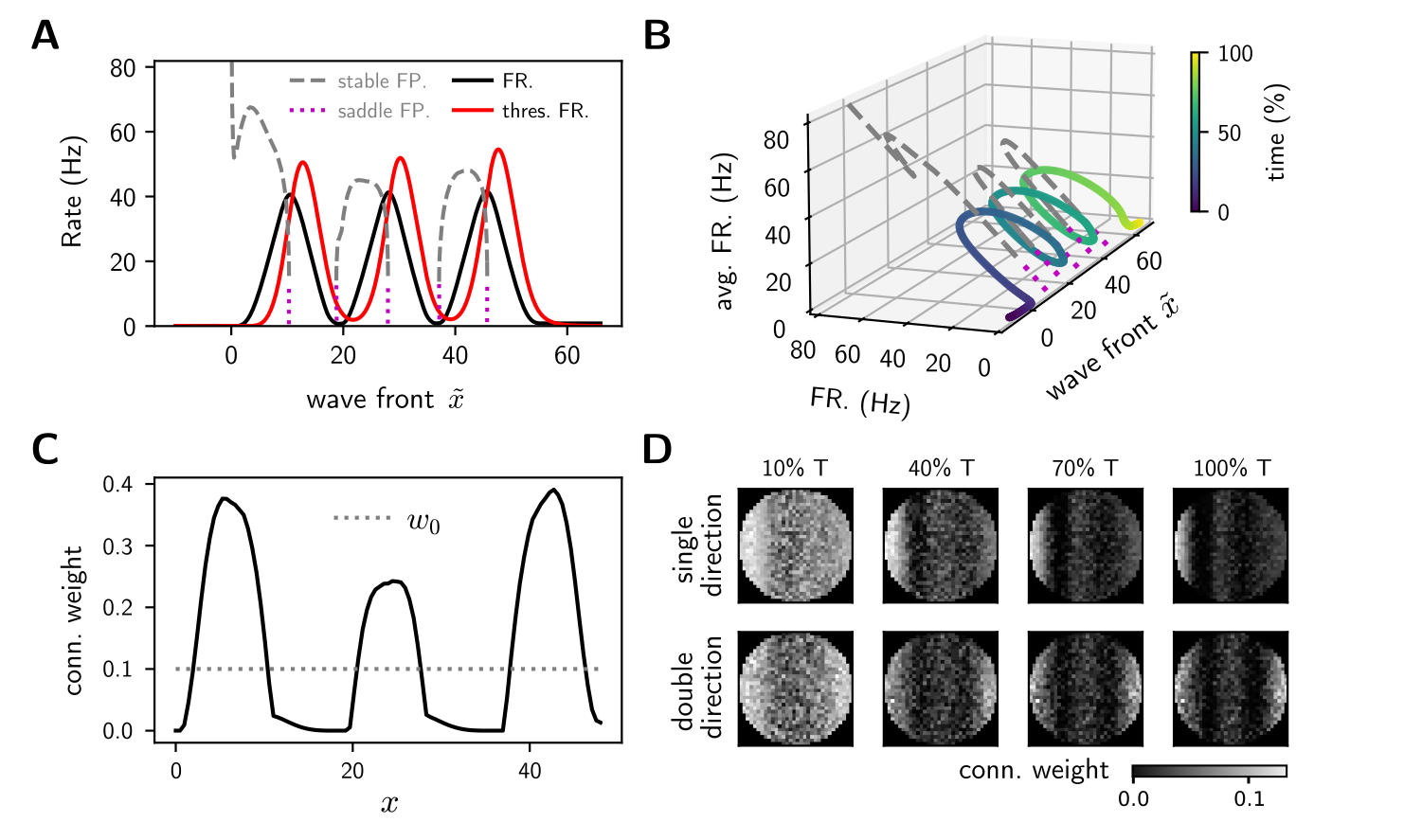}
    \medskip
    \caption{ {\bf Mechanism underlying periodic structure.} {\bf A.} Reduced model performance during the first wave sweep through a larger LGN space ($l=24$), keeping the standard wave width ($d=4$) and wave speed ($v = 4$ deg/s). {\bf B.} In the temporal phase plane (denoted by wave front, $\tilde{x}$), average firing rate and firing rate spiral around the trace fixed points. {\bf C.} Final connection weight distribution shows three peaks, representing a periodic feed-forward RF; initial weight $w_0=0.1$. {\bf D.} The LGN-V1 connection weight profile in the full simulation shows periodic weight distribution over time (averaged over all neurons in a trial). LGN space is $32 \times 32$. The top panel shows wave sweeps of single direction, from left to right, while the bottom panel shows wave sweeps alternating in both directions horizontally. Parameters for reduced model(full simulation): LTD ratio equals $0.3(0.3)$, $w_0 = 0.1(0.03)$}
    \label{fig:periodic_mechanism}
\end{figure}

Expanding upon the mechanism of crossings of the firing rate and its threshold leading to peaks in the connection weight profile, we increase the length of the LGN space in the reduced model to allow for the possibility of more than one peak. Indeed, by extending the LGN space from 16 to 24 in the reduced model, additional crossings of the firing rate and threshold firing rate occurs; see Figure \ref{fig:periodic_mechanism}A. Here, the crossing happens when the stable fixed point annihilates the saddle node as dictated by the dynamics of the reduced model. The firing rate rises with the creation of the fixed points and quickly dips after the annihilation of the fixed points. In the 3-D temporal phase space spanned by the firing rate and its average, we find the system exhibiting a coiled periodic structure (Figure \ref{fig:periodic_mechanism}B). These dynamics lead to the emergence of a third peak in the connection weight along the 1-D LGN pool (Figure \ref{fig:periodic_mechanism}C) and verified in the original simulation; see Figure \ref{fig:periodic_mechanism}D where vertical waves are presented in from left to right in the top row and alternated in both directions in the bottom row. 

\subsection*{Application: Gap junctions amend local continuity of the retinotopic map}
\label{sec:gap_junction-retinotopy}
So far, the V1 neurons in this model have been uncoupled.
While cells in mouse V1 have yet to form cortical synapses during the developmental time of stage II waves, they are sparsely coupled with gap junctions \cite{yu_specific_2009}.
Previous experimental and computational studies have shown that cells coupled by a gap junction during development tend to have more similar RFs at eye opening (in terms of preferred orientation) than those that were not coupled \cite{ko_emergence_2013, crodelle_modeling_2021}.
Here, we introduce gap junctions into the full simulation model and explore their effect combined with stage II waves on local retinotopic map refinement, expecting local gap-junctions during stage II waves will correct visually mismatched V1 cells through their nearest neighbors.

\begin{figure}[ht!]
\centering
    \includegraphics[width=5in]{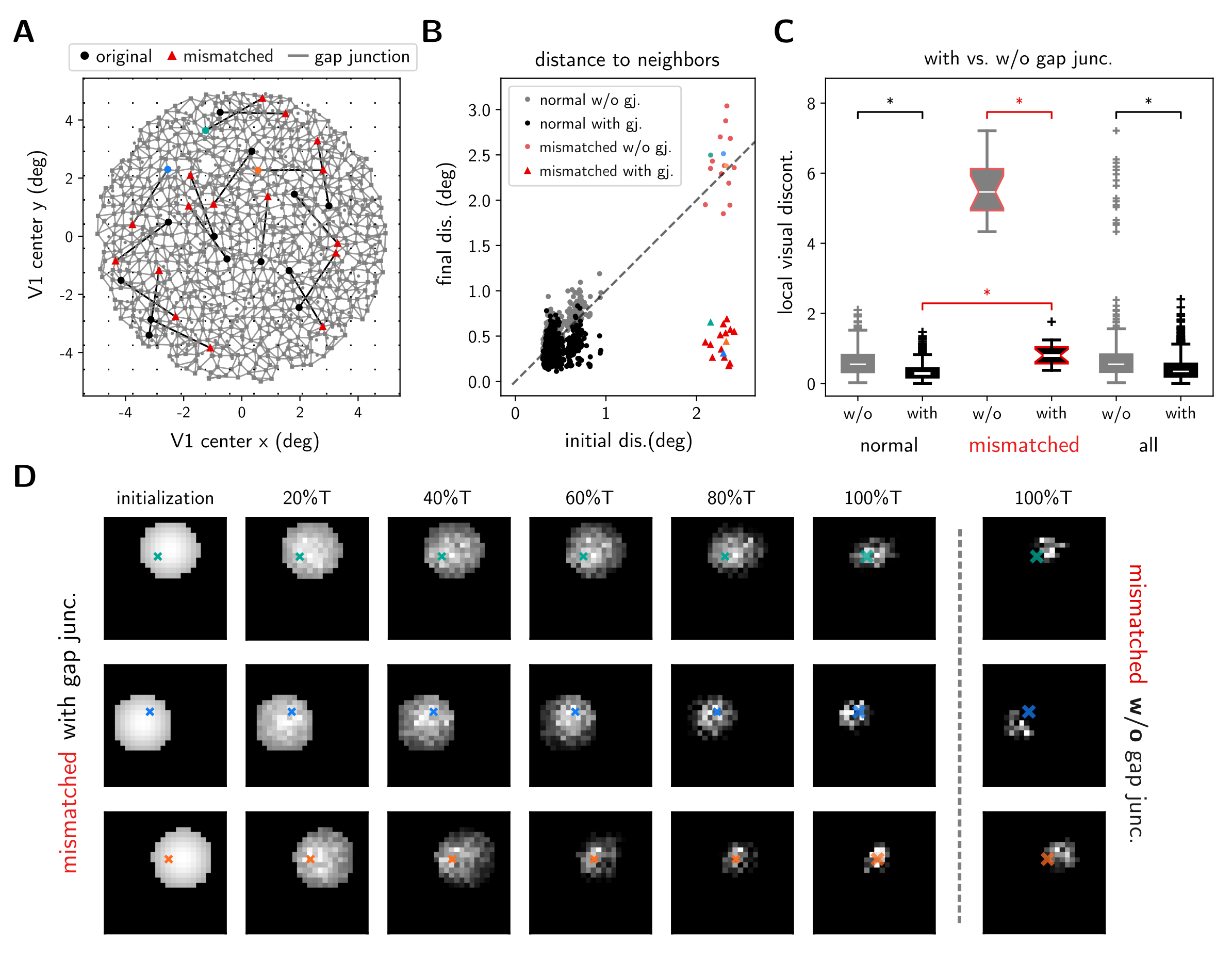}
    \medskip
    \caption{ {\bf Retinotopy refinement with gap junctions.} {\bf A.} Schematic of the connectivity within V1 and its corresponding center in the $24\times24$ LGN grid. Gray squares correspond to V1 cell with normal LGN center and gray lines connecting V1 neurons show gap junction connections to their nearest neighbors. Black dots mark the original RF center of V1 neurons with misplaced RF centers (red triangles). Colored dots match the corresponding sample neurons in B and D. {\bf B.} Scatter plot of the RF center distance to neighbors before (x-axis) and after refinement (y-axis). The gray dots show simulation results without gap junctions, while black dots show results for a network with gap junctions. The dashed line denotes no change in the RF center distance to neighbors with refinement. {\bf C.} Local visual discontinuity (see \nameref{sec:models_and_methods} for definition) is calculated after refinement with and without gap junctions. V1 neurons are grouped into those with normal RF centers, those with misplaced RF centers, and the third group contains all V1 neurons.  {\bf D.} Example refining process of V1 neurons with a misplaced RF center with gap junctions and without (right-most column). The cross marks the original RF center away from the center of the misplaced available LGN pool.}
    \label{fig:retinotopy_refinement}
\end{figure}

In the new setup shown in Figure \ref{fig:retinotopy_refinement}A, the 1024 V1 cells now tile up a circular visual field with scattered visual positions, and 80\% of the V1 cells (gray squares) are allowed to be electrically coupled to their nearest neighbors with a probability of 75\%, i.e., 4-6 gap junctions per neuron (depending on their distance to the boundary). 
While this is a higher coupling percentage than determined experimentally \cite{yu_preferential_2012}, we note that the results hold across changes in the gap junction connection probability (see \nameref{SIfig:GJprobabilities}).
The LGN pool is now extended to 24 $\times$ 24 such that each V1 cell now samples from a circular subset of its local 16 $\times$ 16 subgrid of LGN cells that share the same center as the V1 cell.
Additionally, we randomly place 16 gap-junction coupled V1 cells to sample from an LGN pool (under the red triangles) that does not match with its supposedly visual field center (black circles), in contrast with their neighboring V1 cells.
These V1 cells are referred to as ``mismatched" cells.

We then measure the distance between the RF center of each V1 cell to the averaged RF center of its nearest neighbors before and after the stage II waves (see Figure \ref{fig:retinotopy_refinement}B). 
Consistent with previous studies, gap junction brings coupled V1 cells closer in the visual field (black dots) compared to the same model that block the gap junctions (gray dots), as shown in Figure \ref{fig:retinotopy_refinement}B.
In addition, the centers of the RFs belonging to the ``mismatched" V1 cells (red triangles) were pulled much closer to the center of their gap-junction-coupled nearest neighbors, than those that did not have a gap junction (pink dots).

Using a measure of local visual discontinuity (see \nameref{sec:models_and_methods} for details), we show that, after the stage II waves, the V1 network that includes gap junction couplings has a more continuous local map of RF centers (lower local discontinuity) than a network that did not have gap junctions; see first two box plots in Figure \ref{fig:retinotopy_refinement}C. This difference is much more striking when considering a network with mismatched RF centers; see middle two box plots in Figure \ref{fig:retinotopy_refinement}C.
Although the RF centers of the refined misplaced neurons are still significantly different from the normal ones (red significance bar in Figure \ref{fig:retinotopy_refinement}C), the overall effect of including gap junctions is still clearly significant; see final two box plots in Figure \ref{fig:retinotopy_refinement}C and D.
These results suggest that gap junctions may play a role in smoothing the retinotopic map along with the stage II retinal waves.

\subsection*{Application: Role of stage II waves in seeding orientation preference}
\begin{figure}[ht!]
    \centering
    \includegraphics[width=5in]{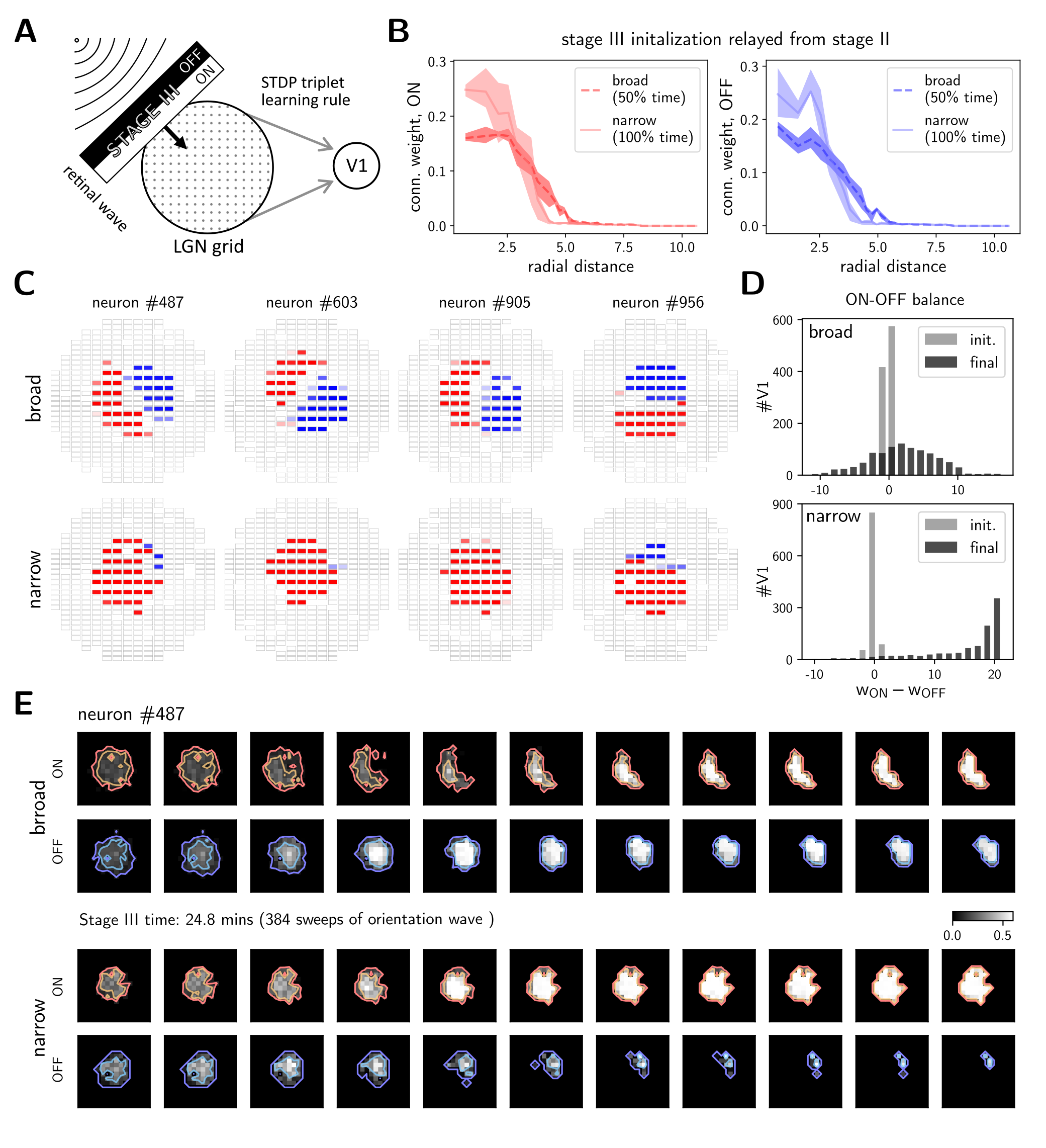}
    \caption{{\bf Influence of refining during stage II waves on formation of orientation selectivity after stage III waves.} Randomly picked trial from standard parameter of stage II retinal waves. {\bf A.} Schematic of the stage III wave presented to the LGN. {\bf B.}  Final distribution of connection weights from ON (left, red) and OFF (right, blue) LGN cells after 50\% (broad) and 100\% (narrow) of the default number of presentations of stage II waves plotted against radial distance. {\bf C.} Distribution of difference between ON and OFF LGN cells for all V1 cells. {\bf D.} In each box the synaptic weight trajectory is shown for the LGN cells whose synapses potentiated from the initial strength for simulations in which the size of the RF after stage II is small (top row) and large (bottom row). Red curves correspond to ON LGN cells and blue to OFF cells. (right) histogram of the difference in connections strength from ON and OFF cells for the small (top) and large (bottom) starting RFs. {\bf E.} Time dynamics for the synaptic weights from LGN to a sample V1 cell for small (top two rows) and large (bottom two rows) starting RFs. Lighter colors correspond to stronger weights. }
    \label{fig:stage3OS}
\end{figure}

Another functional implication of changes in RF structure during stage II waves involves the subsequent refining that may occur during stage III.
Stage III waves, i.e., glutamatergic retinal waves, are faster and more spatially localized than stage II waves and, most notably, recruit ON and OFF retinal cells in succession rather than simultaneously as in stage II waves\cite{kerschensteine_precisely_2008} (illustrated in Figure \ref{fig:stage3OS}A).
It is thought that stage III waves contribute to the formation of orientation preference\cite{kerschensteiner_glutamatergic_2016, thompson_activity-dependent_2017}, a trait of V1 cells where their ON and OFF LGN inputs are spatially offset and aligned to a particular orientation angle \cite{hubel_simpleRF_1959}.

In this section, we demonstrate the formation of orientation preference during stage III wave and how it depends upon the resulted RF relayed from stage II waves. To achieve this, we manually paused the stage II simulation (with the default parameter set) after 50\% or 100\% of the full simulation time, which results in a flat and broad disk-like RF (50\%) and tall and narrow disk-like RF (100\%), respectively (see Figure \ref{fig:stage3OS}B). Note that similarly-shaped RFs may also arise from changes in stage II wave speed and width (recall Figures \ref{fig:wave_speed} and \ref{fig:wave_width}). Then, we replace stage II waves with randomly oriented stage III waves that repeatedly sweeping through LGN cells, activating first ON, and then OFF cells during each wave; the same triplet STDP learning rule is applied. See \nameref{sec:models_and_methods} for detailed setup.

Our simulation shows that if the initialized RF relayed from the stage II has a moderated size and strength (denoted as broad) the resulted RF induced by stage III waves will acquire a stable orientation preference (see Figure \ref{fig:stage3OS}D, top row). 
In this simulation, although the same random sequence of oriented waves is used to stimulate a batch of 1024 neurons, each neuron still acquires its own orientation preference due to the 95\% randomly chosen LGN connection. Orientation biases are seeded by this randomness, and then fixated by the accumulated weight changes resulted from the waves with corresponding orientations (see Figure \ref{fig:stage3OS}E, top row).
The triplet STDP rule always lead the connection to the extremes (i.e., either the max or min weight allowed), just like the pair-wise STDP rule that eventually breaks the balance between LTP and LTD and leads to a bimodal weight distribution\cite{watt_homeostatic_STDP_2010}.

On the other hand, if the relayed RF has a smaller RF with stronger connection strengths (denoted as narrow), then the V1 cells can hardly form any orientation preference at the end of stage III, except for ON-dominated iso-oriented RFs (as shown in Figure \ref{fig:stage3OS}D, bottom row). Due to the stronger connections, stage III waves always lead to more potentiation for the active ON synapses while the OFF synapses near the exiting direction of the wave are depressed, (explained earlier in \nameref{sec:reduced_model}).
As the waves reoccur for each orientation, the potentiated ON synapses now activate the post-synaptic V1 cells more easily, therefore pushing the front of this potentiation toward the incoming direction of the wave (at all orientations), leaving almost all the OFF synapses depressed. 
Such a process is irreversible since the counteracting depression onto these ON synapses are mediated by the competing OFF synapses, which are weaker in strength and small in number due to the smaller RF, as exemplified in Figure \ref{fig:stage3OS}E bottom row. 

Thus, under the assumption of ON-OFF balance (see a discussion on the topic in \nameref{sec:on-off-balance}), our model predicts that when the stage II period persists for too long, or the LGN synapses are pruned too much even with stronger connection strength, the resulted RF will be dominated by the potentiated ON synapses obscuring any initial bias in orientation. 

\section*{Discussion}
In this work, we constructed a spike-timing plasticity model of the LGN-V1 pathway during an early developmental phase in which stage II retinal waves induce ON-OFF synchronized spontaneous activity in the LGN. In this model, we explored the effects of changing wave speed and wave width on the synaptic refinement process underlying the development of disk-like feed-forward RFs in V1, and contrasted it with a ring-like RF shape when increasing initial connection strength or LTD to LTP ratio. 
We further clarified the mechanism underlying these effects through the analysis of a reduced rate-based model. 
With the reduced model, we uncovered the possibility of periodic RF development with the same conditions that lead to ring-like RFs when there is a larger pool of available feed-forward connections, and verified the results in the full spiking model.
Adding known gap junctions to the V1 network, we showed that gap junctions in the presence of stage II retinal waves can help recover local retinotopy even when some V1 neurons' initial visual field centers are mismatched with the local LGN cell.
Finally, extending the developmental period into stage III waves, we found that V1 neurons with broader circular RFs at the end of stage II were more likely to develop orientation selectivity at eye opening, illustrating the importance of stage II RF refinement.
This study clearly identifies the potential mechanisms affecting the feedforward synaptic structure from LGN to V1 during the first two postnatal weeks of development.
These results may be valuable in understanding the developmental processes underlying the formation of synaptic pathways in other cortical areas and during different developmental time periods.

\subsection*{Synaptic pruning and potentiation during development}
Stage II waves are driven by acetylcholine release from starburst amacrine cells \cite{feller_requirement_1996}. Experiments in the superior colliculus (SC) \cite{mclaughlin_retinotopic_2003}, dorsal LGN \cite{grubb_abnormal_2003, pfeiffenberger_ephrin-as_2006} and primary visual cortex \cite{cang_development_2005} show that removing the acetylcholine receptor abolishes the retinal waves \cite{bansal_mice_2000, dhande_development_2011} and results in much coarser retinotopic maps after development. The much larger axonal spread of retinal ganglion cells (RGCs) found in these experiments indicates a synaptic pruning process may be missing in the absence of stage II retinal waves.
Additionally, a very recent study specifically showed that retinal waves instruct RGC axonal pruning on the periphery and growth in the center \cite{matsumoto_hebbian_2024}.
From the perspective of the postsynaptic neuron (V1 cells in our study, SC in experiments), these results are consistent with our model prediction: that stage II waves typically prune synapses in the surround and potentiate those in the center.
Experiments specifically measuring the thalamocortical axonal spread in V1 during development are missing.

\subsection*{Wave speed and width}
Our model results show that size variability in V1 receptive fields \cite{niell_highly_2008, vandenbergh_RFproperties_2010} may be at least partially explained by the variability in wave speed and width of stage II retinal waves \cite{maccione_following_2014}.
While previous experiments have achieved simultaneous changes in the speed and width of the retinal waves using cyclic adenosine monophosphate (cAMP)\cite{stellwagen_dynamics_1999}, they did not investigate the resulting effects in cortex. 
A subsequent modeling study of ocular competition in dLGN with cAMP implied a RF refining effect, which is similar to ours, through a Hebbian burst-based learning rule for retinogeniculate synapses\cite{butts_burst-based_2007}.
However, the underlying mechanisms, are obscured by the simultaneous consideration of changes in width and speed (determined by cAMP), which further influences the level of activation and the time window for learning.
Therefore, to verify the results (or predictions) in our model experimentally, well-controlled manipulation of wave properties and synaptic dynamics are needed, such as spatiotemporally controlled release of acetylcholine in the retina.

\subsection*{Learning rules}
Spike-time-dependent learning rules and its variants have been found in various adult sensory cortices \cite{feldman_barrel_LTPLTD_2000,schnupp_plasticity_A1_2006,sjostrom_visual_stdp_2001} and during development\cite{larsen_stdp_2010} ever since its first discovery in the hippocampus \cite{bi_stdp_1998}. 
In the visual cortex specifically, the triplet learning rule \cite{pfister_triplets_2006} that we applied here has been shown to fit well with experimental results\cite{sjostrom_visual_stdp_2001}.
In addition, the modeling study on a burst-based learning rule found in the retinogeniculate synapses during stage II waves, suggests that modified STDP rules with longer time windows and involving multiple spikes, like the triplet rule, can better explain the observed plasticity over naive STDP rules \cite{butts_burst-based_2007} (but see \cite{bennett_refinement_2015}).
This work observed less RF refinement with cAMP application, which increases the size, speed and frequency of the waves\cite{stellwagen_dynamics_1999}. Those results are closely related to our observation on RF refinement, albeit ours is in V1, the next stage in the early visual pathway. With more carefully controlled numerical experiments, here we demonstrated that a triplet STDP learning rule, reducible to BCM when assumed Poisson \cite{gjorgjieva_triplet_2011,zenke_synaptic_2013}, can refine single neuron RF as well as the local retinotopy, and clarified how the triplet learning rule leads to the various effects with the reduced model analysis.

We also introduced a type of homeostasis that restricts the sum of incoming synaptic weights to one neuron to stay close to its initial value on the timescale of inter-wave interval, similar to synaptic scaling \cite{keck_synaptic_homeostasis_2013} (for details see \nameref{sec:models_and_methods}).
This type of homeostasis encourages competition between synapses, effectively reducing the number of waves needed for observing the RF refinement effect (see Fig. \ref{SIfig:homeostasis_learning_rate}).


\subsection*{Periodic connectivity}
Periodicity in connectivity may arise both within cortical areas or across brain regions. For example, Layer (L)2/3 neurons in primary visual cortices that have orientation columns horizontally connect to neurons across several hyper-columns with similar orientation preference or the same occularity in a periodic fashion through long-range horizontal connections (LHCs) \cite{sincich_oriented_2001, lund_anatomical_2003}. Additionally, grid cells in the medial entorhinal cortex (mEC) have hexagonal RF patterns that map out a subject's movement in a 2-dimensional space \cite{moser_place_2008}, the same structure has now been found in rat S1 and V2 area \cite{long_S1_2021,long_2021_compactV2}. The mechanisms underlying these periodic connections remain elusive. Some experiments from L2/3 V1 neurons suggest that LHCs are independent of retinal activity \cite{ruthazer_role_1996, smith_distributed_2018, gribizis_visual_2019}, while other studies in primary auditory cortex demonstrate an activity-dependent re-wiring of axons \cite{sharma_induction_2000}. In terms of the periodic RF in mEC, studies suggest that the hexagonal RF may result from periodic connections set up during development \cite{naumann_conserved_2016, wang_modularization_2021}; However, periodic spontaneous activity has also been found in L3 of the developing entorhinal cortex \cite{sheroziya_spontaneous_2009}, suggesting a different mechanism. The emergence of periodicity in the RFs in our model relies only on the interaction between an STDP learning rule and retinal wave properties, supporting the case in which periodic connections arise in an activity-dependent way.

A theoretical mathematical model investigating the interaction between 1-dimensional traveling waves and pattern formation found that activity-dependent plasticity can inform periodic RFs \cite{bennett_refinement_2015}. 
However, it is hard to decouple the frequency of multiple incoming waves and the resulting periodicity of the RF in their model.
And they focus more on demonstrating the effects of different STDP rules and noise in the presence of multiple waves than a mechanistic explanation.
In contrast, our model uses biologically-relevant connection profiles, spontaneous retinal wave parameters, and a triplet STDP rule verified for the visual cortex to extract the mechanisms underlying the formation of periodic RFs from decoupled waves of spontaneous activity.

\subsection*{Retinotopic refinement and gap junctions}
Independent of retinal waves, spontaneous activity arises in the cortex during development \cite{khazipov_early_2006, martini_spontaneous_2021}. A recent simplified 1-dimensional theoretical modeling work of retinotopic refinement\cite{wosniack_adaptation_2021} characterizes the interaction of local (L, of thalamic origin) and global (H, endogeneous to cortex) events based on experimental data\cite{siegel_peripheral_2012} and a Hebbian learning rule; the model postulates that the L-events carry retinotopic information while H-events regulate cortical connectivity. 
We note that local recurrent chemical synapses are mostly absent in the cortex during stage II waves\cite{yu_specific_2009}, but cortical activity can still be mediated through gap junctions and silent synapses among subplate neurons\cite{molnar_transient_2020}. 
With gap junctions, we postulate (and verified in our model) that stage II retinal waves are able to refine the local retinotopic map by encouraging nearby neurons to share similar thalamocortical input, even when a few neurons have initially mismatched connections.

\subsubsection*{Gap junctions in V1 during development}
In a more detailed biological picture, thalamocortical synapses are first formed in the subplate and connections to L4 neurons are then guided by these subplate neurons through correlated spiking activity \cite{herrmann_ultrastructural_1994, sengpiel_role_2002, molnar_transient_2020}. During the developmental period coincident with stage II retinal waves (P0-P7), excitatory neurons derived from the same progenitor cell are coupled by gap junctions rather than chemical synapses \cite{yu_specific_2009, colonnese_uncorrelated_2017, molnar_transient_2020}. While the influential role of gap junctions during synaptogenesis in the next developmental stage (when gap junctions are effectively gone during stage III waves) has been deeply explored by experiments and models \cite{ko_emergence_2013, yu_preferential_2012, crodelle_modeling_2021}, little is known about its role during stage II retinal waves.

A very recent experiment on the superior colliculus of the adult mouse has shown that postsynaptic cells have an imprecise retinotopy despite a precise RGC axonal tiling \cite{molotkov_topographic_2023}, consistent with a study on local retinotopy of mouse V1 \cite{smith_parallel_2010}. These results support the existence of spatially-biased random sampling of incoming connections on the single-neuron level in the developing cortex. In our work, we propose that a gap-junction-coupled network, which promotes correlated activity in the nearest neighbors, could help mitigate such bias in single neurons, without which the resulting retinotopy could be even less precise. 

Note that our model effectively considers subplate neurons and immature L4 neurons as the same population, some of which is coupled by gap junctions. In reality, vertically aligned sister cells are the only neurons in this time period that have been shown to have gap junction connections \cite{yu_preferential_2012}. We have shown that our result -- that gap junction connections between neurons during stage II retinal waves serve to refine and correct the local topographic map -- still holds across different percentages of gap-junction coupling (see Fig. \ref{SIfig:GJprobabilities}), even when that coupling is reduced to only 1-2 connections per neuron (as in the developing cortex). 

\subsection*{Orientation preference and stage III retinal waves}
In contrast to ON-OFF synchronous stage II retinal waves, stage III waves are known to be ON-OFF asynchronous in the RGCs and LGN cells, where the OFF cells are activated right after the neighboring ON cells within each wave \cite{kerschensteine_precisely_2008}. Experimental evidence suggests that orientation selectivity is formed by spontaneous activity before eye-opening (around P14), since broad tuning of V1 neurons can be observed at eye-opening \cite{ko_emergence_2013, hoy_layer-specific_2015}. The asynchronous activation during stage III waves makes it a perfect candidate for early development of orientation preference in the primary visual cortex \cite{kerschensteiner_glutamatergic_2016, thompson_activity-dependent_2017}, and our preliminary modeling of the stage III waves with triplet STDP rule indeed supports this hypothesis. As the STDP triplet rule can be reduced to a BCM-like rule in a rate-model\cite{}, our result is also consistent with an earlier study that shows the segregation of ON-OFF subregion is dependent on BCM rule operating in a linear region using natural images \cite{lee_BCM_LGN_2000}.

In addition, our model shows that the condition for the emergence of orientation selectivity also depends on the feed-forward connection profile (or RF) initialized during stage II retinal waves. A narrow bell-shaped RF with little room for non-overlapped ON and OFF LGN connections can make it very hard to develop orientation selectivity, demonstrating the importance of the width and speed of individual stage II waves for proper RF development in later stages. These results may help explain the high variability in the selectivity of V1 neurons.
\subsubsection*{ON-OFF subregion imbalance}
\label{sec:on-off-balance}
Our model would predict that neurons with narrow RF at the end of stage II will eventually acquire a ON-dominated RF at the end of stage III.
However, this prediction relies on the assumption that the ON and OFF inputs are balanced in each V1 cell. 
We note that extensive experimental studies have shown that OFF RGC and OFF LGN cells are more active than ON cells\cite{ratliff_retina_2010,jiang_asymmetry_2015}, due to their differences in the numbers and anatomical pathways\cite{ratliff_retina_2010, ichinose_pathways_2022}.
For simplicity, we drop this asymmetry between the ON and OFF synapses in our model, whereas in reality, the intrinsic bias toward OFF cells may give enough room for the emergence of orientation preference int the visual system, tolerating an extended range of parameters in the stage II retinal waves \cite{maccione_following_2014}.

\section*{Conclusion}
In summary, we have used two levels of mathematical modeling to identify and demonstrate potential mechanisms underlying the refined RFs of V1 cells as a function of different characteristics of the retinal waves. Our model predicts that periodic RFs may arise due to changes in long-term depression or initial connection strengths, gap junctions in V1 during the first postnatal week may play a role in increasing local visual continuity, and variability in the shape and size of the RF at the end of stage II influences the emergence of orientation preference of V1 cells during stage III retinal waves (before eye-opening). These results are not only useful in understanding activity-dependent development of V1 RFs, but can also be used for predictions across different sensory cortices during development. 

\section*{Models and Methods} 
\label{sec:models_and_methods}

\subsection*{Setup of the developmental model}
This model describes feed-forward learning of V1 receptive fields in the developmental stage, driven by successive random spontaneous stage II waves originating in retinal ganglion cells (RGCs) and relayed by LGN cells to V1. We model 1024 local V1 neurons that share the same visual field with a standard 16$\times$16 grid of LGN cells. Each V1 neuron is post-synaptic to 80\% of the LGN cells, randomly chosen, in the grid initially. The network we consider is feed-forward only, i.e., V1 lacks intracortical chemical synapses during this stage \cite{li_synaptogenesis_2010}. They do have gap-junction connections, which we add to the model later \cite{li_clonally_2012}.

\subsubsection*{Stage II waves}
Many more retinogeniculate synapses (around 30) per thalamocortical projecting cell exist in dLGN during development than a mature one (around 10, but functionally 1-3)\cite{litvina_functiona_2017a}. This indicates an abundance in the local activation of dLGN from retinal waves. Thus, we assume that LGN activity is directly proportional to the amplitude of the retinal waves, i.e. we consider a simplified linear mapping from retina to LGN. Focusing on a local population of LGN cells, we model the effect of the retinal waves as drifting sinusoidal bars coming from random directions, mimicking parallel wave fronts. 

This is a valid assumption when the distance from the origin of these waves is large compared to the size of the local population.
According to the thorough investigation of retinal wave characteristics by Maccione et. al. in 2014\cite{maccione_following_2014}, stage II retinal waves travel around 120 {\textmu}m/s. 
In our model, this translates to around 4 degrees/s or 3.2 units of distance between the nearest LGN cells per second. Stage II waves have directional bias dependent on different quadrants but are still largely random \cite{stafford_spatial_2009} so we only apply an uniform distribution of wave direction in our model. Assuming balanced ON-OFF coverage of a local visual field, for simplicity, we construct two grids with overlapping vertices of the same size for ON and OFF LGN cells. The standard stage II wave width is 10 degrees, or 8 in the unit of interval between LGN cells.

\subsubsection*{Stage III waves}
Our modeling of the stage III waves follows the simplicity of stage II waves as oriented drifting bars activating the LGN cells, but with OFF cell activation lagging behind the ON cells in each sweep instead of overlapping with each other (see Fig \ref{fig:stage3OS}A), though the total width of the wave is still set as 10 degrees.
However to reflect other difference in the experimental observations, the speed of stage III wave is faster, 5 degree/s, and each sweep is repeated three times (\cite{kerschensteine_precisely_2008,maccione_following_2014}).
Also see \nameref{tbl:waves} for a table of the standard parameters.

\subsubsection*{Models of V1 neurons and LGN cells}
In the full simulations, V1 neurons are modeled by conductance-based adaptive exponential integrate-and-fire models\cite{brette_adaptive_2005}:
\begin{eqnarray}
    \label{eqn:EIF}
    \left\{\:
    \begin{aligned}
        C_{m}\frac{\dd V}{\dd t} &= -g_{L}V_{L} + g_{L}\exp\left[-\frac{V-V_{T}}{\Delta_{T}}\right] + g_{E}\left(V-V_{E}\right) - Q + g_{\text{gap}}\sum_{i}{\left(V - V_{i}\right)}\\
        \tau_{d}\frac{\dd g_{E}}{\dd t} &= -g_{E} + h_{E} \\
        \tau_{r}\frac{\dd h_{E}}{\dd t} &= -h_{E} +  \sum_{j}w_{j}\sum_{k}\delta\left(t-{t^{\left(j\right)}_{k}}\right) \\
        \tau_{Q}\frac{\dd Q}{\dd t} &= a\left(V-V_{L}\right) - Q
    \end{aligned}, \right.\kern-\nulldelimiterspace
\end{eqnarray}
where $V$ is the membrane potential and $Q$ is the adaptive current. When $V$ surpasses the firing threshold, $V_T=-50$mV, it exponentially increases until it reaches $0$mV, whereupon it gets reset to $V_{r}=-65$mV. At the time of reset, $Q$ is increased by $b = 2.5$pA. The synaptic conductance, $g_{E}$, comes from the feed-forward spike train $t^{\left(j\right)}_{k}$, containing $k$ many spikes from connected LGN cells indexed by $j$. For those simulations that contain gap junctions, $g_{\text{gap}} = 0.75$nS, otherwise it is set to 0. Also see Table \ref{tbl:EIF}. 

LGN cells are modeled by a linear-nonlinear-Poisson (LNP) process with parameters extracted from experiments on mouse LGN\cite{grubb_quantitative_2003}. 
When modeling the developmental process in the absence of visual experience (during first two postnatal weeks when eyes are closed), the linear convolution in the LNP is directly replaced by a normalized retinal wave amplitude, $I\in\left[0,1\right]$. The nonlinear (sigmoid) gain function for the LGN firing rate, $\hat{r}$, takes the form of a generalized logistic function:
\begin{eqnarray}
    \hat{r}\left(I\right) = A + \frac{B}{1+\exp\left[K\left(c_{50}-I\right)\right]},
\end{eqnarray}
where parameters $A$ and $B$ can be found by solving $\hat{r}(0)= 3Hz$ (spontaneous firing rate) and $\hat{r}(1) = 60Hz$ (maximum firing rate). The strongest gain occurs for $K=3$, and $c_{50} = 0.25$ marks the input strength that results in half of the maximum output. From this process, a time-dependent Poisson spike train is generated.

\subsubsection*{Learning rule of the LGN-V1 connections}
We apply the STDP triplet rule as the learning rule in our model, which is suitable for describing plasticity in the visual cortex\cite{pfister_triplets_2006}.
By employing one pre-synaptic tracer variable $\zeta_{\scriptscriptstyle{-}}$ and two post-synaptic tracer variables $\zeta_{\scriptscriptstyle{+}}$ and $\zeta_{\scriptscriptstyle{slow}}$, the weight change $\Delta{w}$, can be expressed as the following:
\begin{eqnarray}
    \label{eqn:triplet}
    \Delta{w} = 
    \begin{cases}
        A^{\scriptscriptstyle{+}}\zeta_{\scriptscriptstyle{+}}(t)\zeta_{\scriptscriptstyle{slow}}(t-\epsilon), &\text{if }\:t = t_{post}\\
        A^{\scriptscriptstyle{-}}(t)\zeta_{\scriptscriptstyle{-}}(t), &\text{if }\:t = t_{pre}\\
        0, &\text{otherwise}.
    \end{cases}    
\end{eqnarray}
The synaptic tracer variables $\zeta_{k}$, ($k =\:\scriptstyle{-,+,slow}$) increment by 1 at the time of a spike in their corresponding pre-synaptic neuron $t_{pre}$ or  post-synaptic neuron $t_{post}$, and  otherwise decay exponentially. 
Additionally, we add a homeostatic fast-rate detector introduced by Zenke et. al., 2013\cite{zenke_synaptic_2013} to balance LTP and LTD by setting a target firing rate $r_{0}$ for V1 neurons as follows:
\begin{eqnarray}
    A^{\scriptscriptstyle{-}} = \frac{\tau_{\scriptscriptstyle{+}}\tau_{\scriptscriptstyle{slow}}{[\olsi{r}(t)]}^2}{\tau_{\scriptscriptstyle{-}}r_{0}}A^{\scriptscriptstyle{+}},
\end{eqnarray}
where the average firing rate of the V1 neuron is $\displaystyle \olsi{r} = \frac{1}{\tau}\sum_i\exp\left(-\frac{t-t_{i}}{\tau}\right)$. Parameters for the STDP rule can be found in \nameref{tbl:plasticity}.
\subsubsection*{Optional Homeostasis}
In the full simulations we applied an optional slow homeostasis onto the synapses to enhance the competition between LTP and LTD and shorten the time for simulation apart from the homeostatic fast-rate detector. In the mechanistic analysis of reduced model, we intentionally turn off this homeostatic change to clearly demonstrate the effects, except Fig.\ref{fig:mechanism}G H and I where it is retained to compare with the full simulation. 
The slow homeostasis gradually relax the total synaptic weight of a V1 neuron back to its initial value on the time scale of 2.5s, comparable with the interval between two stage II waves. The dynamics of the total synaptic weight follows a simple ODE:
\begin{equation}
\tau_h\frac{\text{d}s}{\text{d}t} = -(s - s_0),
\label{eqn:homeostasis}
\end{equation}
where $s$ is the total synaptic weight for each neuron and $s_0$ is its initialized value. For each time step, we add the total change in synaptic weights $\sum_i \Delta w_i$ from LTD and LTP to $s$ and evolve the dynamics to get the total change $\Delta s$ due to this homeostasis, and then redistribute the change equally to each individual synapse.
For detailed comparison of this homeostatic effect see Supplementary Fig.\ref{SIfig:homeostasis_learning_rate}.

\subsection*{Receptive field analysis}
In this work, the difference between ON and OFF LGN cells are largely ignored because ON-OFF segregation is not present during stage II retinal waves \cite{huberman_mechanisms_2008}. Therefore, we derive metrics that characterize the size and overall shape of the feed-forward connections by which we define the receptive field (RF), ignoring ON and OFF designations.
The metric we call ``weighted radius'' is used to quantify the spread of the disk-like RF for each V1 neuron and is defined as follows
\begin{equation}
    R = \sum_{k,l}{w_{kl}\left|\vec{x}_{kl} - \frac{1}{N}\sum_{i,j}{w_{ij}\vec{x}_{ij}}\right|},
    \label{eqn:weighted_radius}
\end{equation}
where $w_{ij}$ is the weight of the connection from the LGN cell at location $\vec{x}_{ij} = (i,j)$ in the visual field.
To quantify the actual amount of connections with enhanced strength in the same unit as weighted radius, we take the square root of the area of increased connection strengths (compared to their initial value) and divide it by two, called the characteristic length of the RF.
The characteristic length for a 1-dimensional RF, as in the reduced model, is defined as the width of the weight distribution at the initial connection strength (i.e., the difference between the red dots in Figure \ref{fig:mechanism}G and H) divided by two.

In addition to the two metrics above, we construct the radial distribution of connection weights by averaging connection weights along the radial axis to give the most intuitive description of RF regardless of the disk-like or ring-like structure. 

To address the retinotopic continuity of the visual field coverage by the RFs of neighboring neurons in the gap junction study, we define a local discontinuity metric. Assuming the V1 neurons are densely packed in a 2D cortical surface (Poisson disk sampling algorithm is used to generate the physical positions), we quantify ``local" by the optimal distance between those densely-packed neurons, $h$. Then we can obtain a metric that describes the smoothness in the local retinotopic mapping, $D$ for each neuron $i$ as follows:
\begin{eqnarray}
    D_{i} = \left|\vec{p}_{i} - \frac{1}{m}\mathlarger{\mathlarger{\sum}}_{\substack{j\in\left\{\left|\vec{q}_{j}-\vec{q}_{i}\right|<d\right\}}}{\vec{p}_{j}}\right|/h,
\end{eqnarray}
where $\vec{p}$ denotes the position vector for each neuron's RF center in the visual field and $\vec{q}$ denotes the physical location for each neuron in the cortical surface. The ``local" visual center is summed over the $m$ neighboring neurons within distance $h$. 
Values above $1$ indicate considerable discontinuity at the corresponding visual position.


\section*{Acknowledgments}
The authors want to acknowledge Prof. David W. McLaughlin for frequent discussions on the topic which improved the quality of this work.
This work is funded by the Science and Technology Innovation
2030 —— Brain Science and Brain-Inspired Intelligence Project No. 2021ZD0201301 (Wei P. Dai),
the National Natural Science Foundation of China No. 12201125 (Wei P. Dai) and the Science and
Technology Committee of Shanghai Municipality No. 22YF1403300 (Wei P. Dai).


\bibliography{stage_II_wave_generic_template}

\newcounter{offset}
\setcounter{offset}{\value{figure}}
\renewcommand{\thefigure}{S\the\numexpr\value{figure}-\value{offset}\relax}

\renewcommand{\thetable}{S\the\numexpr\value{table}\relax}

\newpage
\section*{Supporting information}

\begin{table}[!h]
\caption{{\bf Default parameter set for the retinal waves.}}
\begin{tabular}{|l|c|c|}
			\hline
			 & {\bf Stage II} & {\bf Stage III} \\ \hline
			Wave speed (deg/s) & 4 & 5 \\ \hline
			Wave width (deg) & 10 & 5 \\ \hline
			Exposed LGN cells (diameter) & 16 & 16 \\ \hline
			Inter-wave interval (s) & 6 & 3 \\ \hline
 			On-Off segregation & False & True \\ \hline			
\end{tabular}
\label{tbl:waves}
\end{table}
\begin{table}[!ht]
\caption{{\bf Parameters for exponential Integrate-and-Fire neuron}}
\begin{tabular} {|l|c|c|c|c|}
   \hline
   Voltage &$V_T$ &$V_{\theta}$ &$V_E$ &$\Delta_T$ \\ \hline
   Value (mV) & -50.0 &-20 &0 & 1.5 \\ \hline
\end{tabular}
\vskip 10pt
\begin{tabular} {|l|c|c|c|c|}
   \hline
   Timescale &$C_m/g_L$ &$\tau_d$ &$\tau_r$ &$\tau_Q$ \\ \hline
   Value (ms) &20 &3 &1 &15 \\ \hline
\end{tabular}
\vskip 10pt
\begin{tabular} {|l|c|c|}
    \hline
    Conductance &$a$ &$g_{\text{gap}}$*\\ \hline
    Value (nS) &0.2 & 0.75 \\ \hline
\end{tabular}
\medskip
\caption*{ *$g_{\text{gap}}$ equals zero except in Section \nameref{sec:gap_junction-retinotopy}}
\label{tbl:EIF}
\end{table}
\vskip 10pt
\begin{table}[!ht]
\caption{{\bf Parameters for synaptic plasticity}}
\begin{tabular} {|l|c|c|c|c|c|c|c|c|}
   \hline
   Parameter & $A^+$ &$r_{\text{LTD}}$ &$\tau_{+}$ &$\tau_{-}$ &$\tau_{\text{slow}}$ &$r_0$ &$\tau$ &$\tau_h$ \\ \hline
   Value & 0.003 & 0.3 & 17ms & 34ms &114ms & 1000ms &6Hz &2500ms \\ \hline
\end{tabular}
\label{tbl:plasticity}
\end{table}
\clearpage
\begin{figure}[t!]
    \centering
    \includegraphics[width=0.8\linewidth]{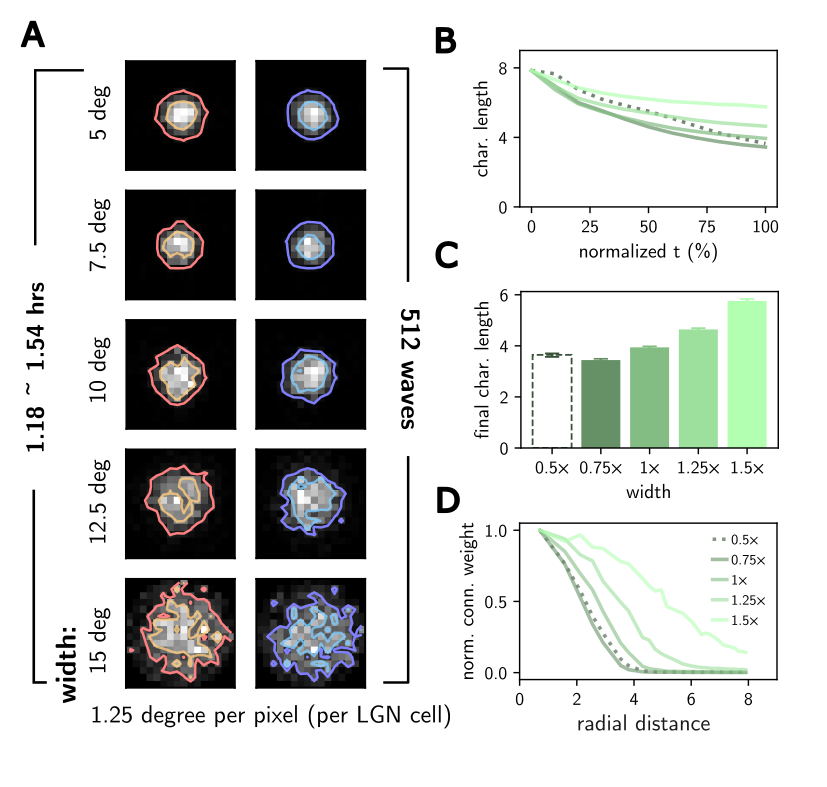}
    \caption{\textbf{The effect of different wave widths but with the same connection strength.}
        \textbf{A.} Two population-averaged LGN-V1 connection profile at the end of stage II with different wave width from a sample trial, labeled with half wave-width in degrees.
        \textbf{B.} Temporal evolution of population averaged characteristic length (CL) over 10 trials with different wave widths.
        \textbf{C.} Errorbars of CL collected from 10 trials over different wave width at the end of stage II.
        \textbf{D.} Radial distribution of connection strength averaged from the 10 trials over different wave widths at the end of stage II.
        In this setup, we use $w=0.3$ for all wave widths, resulting in an insufficient V1 activation for the wave with 5-degree half-width (0.5$\times$ the standard, denoted by dotted lines/contours) such that it does not follow the trend as in the main text. A stronger connection strength will lead to ring structure in the 15-degree wave.
    }
    \label{SIfig:width_same_strength}
\end{figure}

\clearpage
\begin{figure}[t!]
    \centering
    \includegraphics[width=0.8\linewidth]{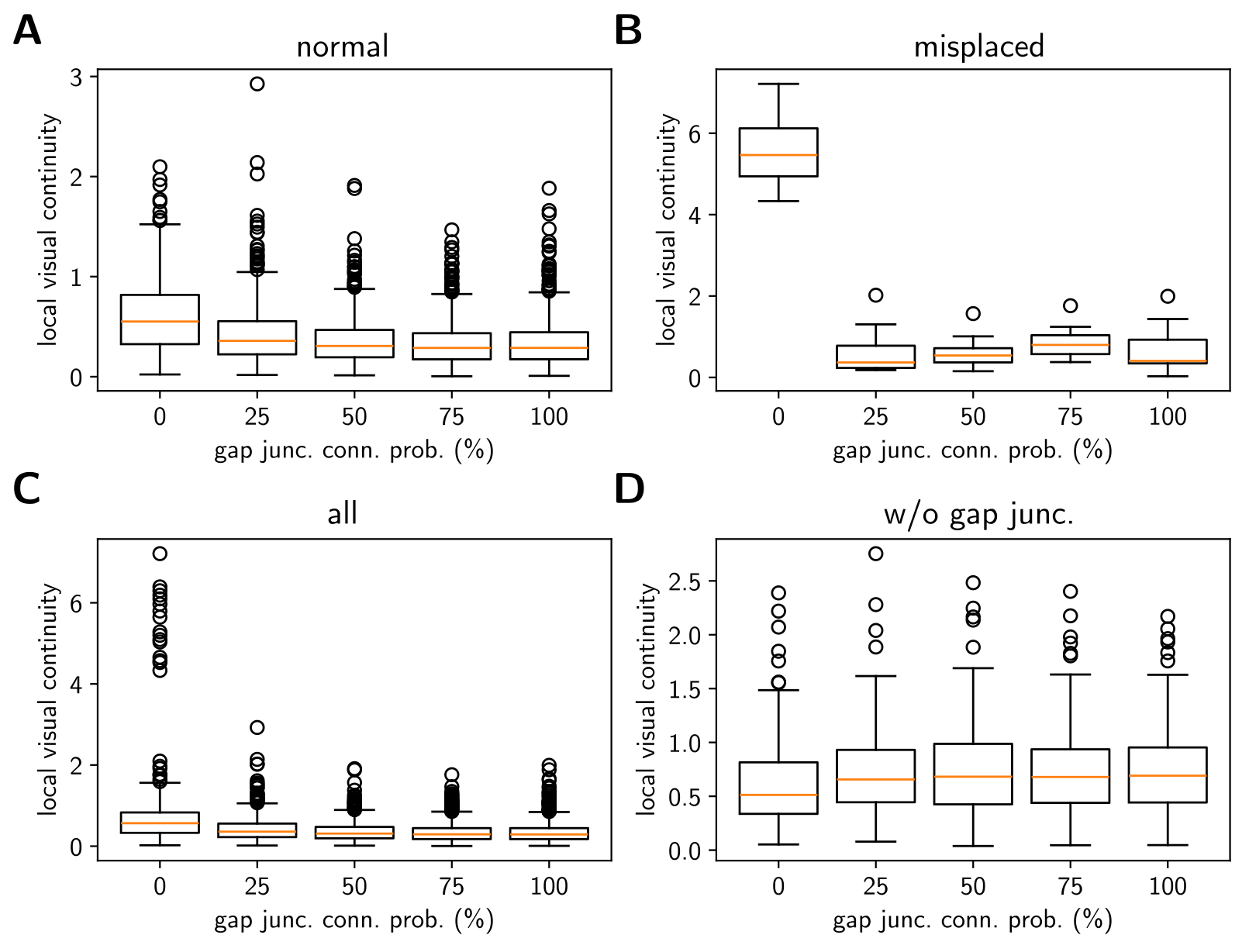}
    \caption{\textbf{Visual continuity for different gap junctions probability}
        Percentages are for the gap-junction-eligible neurons only. I.e., 100\% coupling probability still leaves 25\% of neurons that do not couple with gap junctions. 
        \textbf{A.} Local visual continuity calculated for a gap junction coupled network without misplacement of LGN connections.
        \textbf{B}, \textbf{C} and \textbf{D}. Same metric calculated for misplaced neurons, all neurons and neurons that not eligible to connect with gap junction in the misplaced setup, respectively.
    }
    \label{SIfig:GJprobabilities}
\end{figure}

\clearpage
\begin{figure}[t!]
    \centering
    \includegraphics[width=1.0\linewidth]{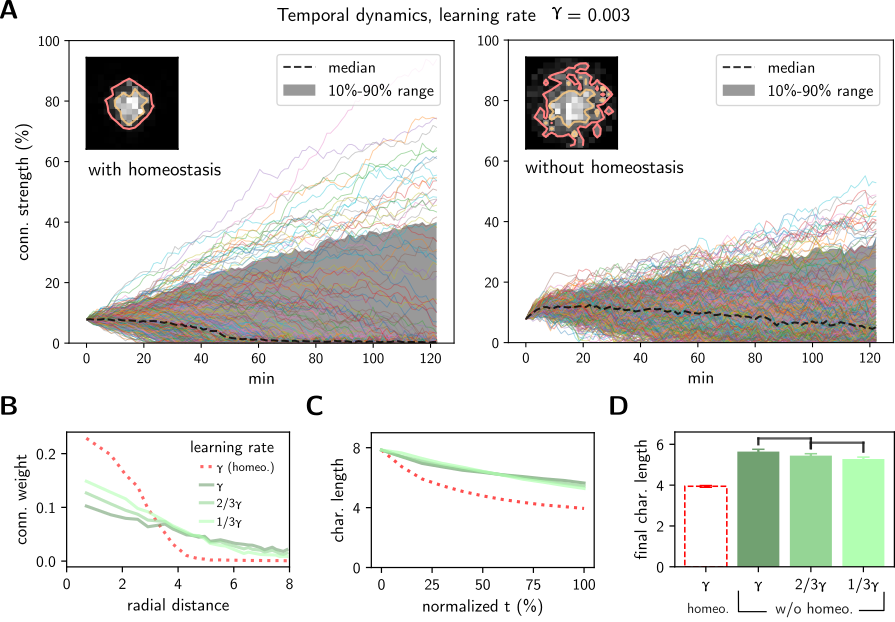}
    \caption{\textbf{Effect of global homeostasis and learning rate on synaptic pruning.} \\
        \textbf{A.} Compare temporal dynamics of the full simulation with (left panel) vs. without global homeostasis (right panel) of the connection strength of a sample V1 neuron, with standard learning rate $\gamma=0.003$. The insets illustrate the final connection profile.
        \textbf{B.} Radial distribution of connection strength averaged from 5 trials for 3 different learning rates without global homeostasis (greens), running time are extended correspondingly (e.g., 3 times for 1/3 learning rate); standard learning rate with global homeostasis is plotted in red for comparison.
        \textbf{C.} Same as B, but for the characteristic length over time.
        \textbf{D.} Barplot for the final characteristic length in B, T-test performed between different learning rates without global homeostasis (T-value = 0.0195 between $\gamma$ and $2/3\gamma$; T-value = 0.0247 between $2/3\gamma$ and $1/3\gamma$).
        Global Homeostasis enhance the competition between synapses, resulting in a faster pruning and a stronger nonlinear structure in the spatial profile.
    }
    \label{SIfig:homeostasis_learning_rate}
\end{figure}

\clearpage
\paragraph*{Appendix. A}
    \label{appendix:A}
    { \bf Derivation of the reduced rate-based model for stage II simulation.}
    For further analysis on the process of stage II wave influence on the thalamocortical connections, we approximate the full simulation of Eq.(\ref{eqn:triplet}) with a rate-based model using a similar set of steps as has been done previously \cite{gjorgjieva_triplet_2011, zenke_synaptic_2013}. First, we reduce the two-dimensional, discrete pool of LGN cells to a one-dimensional, continuous distribution of LGN cells along the $x$-dimension. We assume that increasing $x$ is in the direction of an incoming wave (recall Figure \ref{fig:reducedModel_intro}A). Then, assuming a synapse exists as each $x$ location in the pool, we can write the following equation describing the change in the synaptic weight $\hat{w}\left(x,t\right)$ at time $t$ and location $x$ under the triplet rule\cite{gjorgjieva_triplet_2011}:
    \begin{equation}
    \label{dwdt}
        \frac{\partial \hat{w}\left(x,t\right)}{\partial t} = A^+ z_{\scriptscriptstyle{+}}(x,t) z_{\scriptscriptstyle{\text{slow}}}(t-\epsilon) S^{\text{V1}}(t) -  A^{\scriptscriptstyle{-}}(t)z_{\scriptscriptstyle{-}}(t)S^{\text{LGN}}(x,t),
     \end{equation}
    where $S^{\text{LGN}}(x,t) = \sum_i \delta(t-t_{i}^{x})$ is the spike train with spike times $t_{i}^{x}$ for the LGN cell at location $x$. The spike train for the V1 neuron, $S^{\text{V1}}(t)$, is similarly defined. The pre-synaptic tracer variable $z_{+}(x,t)$ satisfies the differential equation
    \begin{equation}
    \frac{\partial z_{\scriptscriptstyle{+}}(x,t)}{\partial t} = -\frac{z_{\scriptscriptstyle{+}}(x,t)}{\tau_{\scriptscriptstyle{+}}} + S^{\text{LGN}}(x,t).
    \end{equation}
    The post-synaptic tracer variables, $z_{\scriptscriptstyle{-}}(t)$ and $z_{\scriptscriptstyle{\text{slow}}}(t)$, are similarly defined. The LTP and LTD learning rates, $A^{\scriptscriptstyle{+}}$ and $A^{\scriptscriptstyle{-}}$, respectively, are related by a fast-rate detector in the following equation\cite{zenke_synaptic_2013}
        \begin{equation}
        \label{learningRates}
A^{\scriptscriptstyle{-}}(t) = \frac{\tau_{\scriptscriptstyle{+}}\tau_{\scriptscriptstyle{\text{slow}}}{[\olsi{r}(t)]}^2}{\tau_{\scriptscriptstyle{-}}r_{0}}A^{\scriptscriptstyle{+}},
    \end{equation}
    where $r_0$ is the target firing rate and $\olsi{r}(t)$ is the average firing rate of the V1 cell, defined by the following differential equation
       \begin{eqnarray}
        \tau\frac{\dd \olsi{r}}{\dd t} = -\olsi{r} + S^{\text{V1}}(t).
    \end{eqnarray}
We take the expectation over many spike trains and arrive at a rate-based approximation for V1 firing
 \begin{eqnarray}
    \label{eqn:rbar_meanfield}
        \tau\frac{\dd \olsi{r}(t)}{\dd t} & = -\olsi{r} + r.
    \end{eqnarray}
 Then, plug in for $A^{\scriptscriptstyle{-}}(t)$ using Eq.~(\ref{learningRates}) and take expectation over many pairs of LGN and V1 spike trains we write down the rate-based approximation for the change in synaptic weight as
     \begin{align}
          \left<\frac{\partial \hat{w}}{\partial t}\right> &= A^{\scriptscriptstyle{+}} \tau_{\scriptscriptstyle{+}} r^{\text{LGN}}(x,t) \tau_{\scriptscriptstyle{\text{slow}}} [r(t)]^2 -  \left( \frac{\tau_{\scriptscriptstyle{+}}\tau_{\scriptscriptstyle{\text{slow}}}{[\olsi{r}(t)]}^2}{\tau_{\scriptscriptstyle{-}}r_{0}}A^{\scriptscriptstyle{+}} \right)  \tau_{\scriptscriptstyle{-}}r(t)r^{\text{LGN}}(x,t) \nonumber\\
          &= A^{\scriptscriptstyle{+}} \tau_{\scriptscriptstyle{+}}\tau_{\scriptscriptstyle{\text{slow}}} r^{\text{LGN}}(x,t) r(t) \left[  r(t) - \frac{[\olsi{r}(t)]^2}{r_0} \right],\label{eqn:w_partial_meanfield}
      \end{align}   
     where $r(t)$ is the firing rate of the V1 cell at time $t$ and $r^{\text{LGN}}(x,t)$ is the firing rate of the continuous LGN at location $x$ and time $t$. We define the rate-based approximation of connection weight in our model, $w(x,t)$ to be the integral of the approximated change in average, $\left<\partial\hat{w}/\partial t\right>$, from $t=0$ plus some initial value $w(x,0)$. The weight dynamics is BCM-like as it involves a threshold which upon crossing changes the sign of weight change.

Now, we define the firing rate of the V1 cell as the integration of weighted LGN firing over the area activated by the retinal wave, as follows:
    \begin{eqnarray}
    \label{eqn:r_meanfield}
        r(t) = g_0\:\hat{r}\int_{\max(0,vt-2d)}^{\min(2l,vt)}g(x)\:w(x,t)\dd x,
    \end{eqnarray}
    where $g_0$ is the intrinsic gain of the V1 neuron, $l$ is the radius of the pool of LGN cells, and $d$ is the half-width of the wave in units of number of LGN cells (recall Figure \ref{fig:mechanism}A). The function $g\left(x\right) = 2\sqrt{l^2-\left(x-l\right)^2}$ gives the length of a chord in the circular pool of LGN synapses and represents the number of LGN synapses in the vertical direction that are activated at each location of $x$ assuming that they all respond identically to the wave.
    
  Next, we take the time derivative of Eq.(\ref{eqn:r_meanfield}). To make the computation simpler, let the upper limit of integration be $u(t) = \min({2l,vt})$, the bottom limit be $b(t) = \max(0,vt-2d)$ and suppose $Q(x,t) = \int g(x) w(x,t) \dd x$ is the anti-derivative. Then, we can compute the derivative of the integral on the right-hand side of Eq.~(\ref{eqn:r_meanfield}) as follows
    \begin{eqnarray}
        \label{eqn:rt_rhs}
        \frac{\dd }{\dd t}\left(\int_{b(t)}^{u(t)}g(x)\:w(x,t)\dd x\right) = \frac{\dd }{\dd t} \left[ Q\left(u(t),t\right)-Q\left(b(t),t\right)\right].
    \end{eqnarray}
    By the chain rule, the time derivative of $Q(x(t),t)$ gives:
    \begin{eqnarray}
        \begin{aligned}
            \frac{\dd Q(x(t),t)}{\dd t} &= \frac{\partial{Q(x,t)}}{\partial{x}}\:\frac{\dd x}{\dd t} + \frac{\partial{Q(x,t)}}{\partial{t}}\\
            &= g(x)\:w(x,t)\:\frac{\dd x}{\dd t} +
            \int g(x)\:\frac{\partial{w(x,t)}}{\partial{t}}\dd x.
        \end{aligned}
        \nonumber
    \end{eqnarray}
    We again consider a mean-field approximation of the firing rate by taking the expectation over many spikes to write 
     \begin{equation}
     \frac{\dd r}{\dd t}  =  g_0\hat{r} \left[ g(x)w(x,t)\frac{\dd x}{dt}\Big|_{x= b(t)}^{x=u(t)} + \frac{\partial w}{\partial t} \int_{b(t)}^{u(t)} g(x) \ \dd x \right],
     \label{eqn:drdt_long}
     \end{equation}
noting the assumption that the firing rate of the LGN neural field is the constant $r^{\text{LGN}}(x,t)= \hat{r}$ when activated by a retinal wave (inside the bounds of integration) and vanishes otherwise allows us to pull the $\partial w/\partial t$ term outside the integral.  
     
We make a final simplifying assumption to reduce this model to a 2-dimensional system of $r(t)$ and $\olsi{r}(t)$.
To do this, we assume that the learning rate, $A^{\scriptscriptstyle{+}}$, is small enough such that the weight, $w(x,t)$, during a wave can be well-approximated by the weight at the start of the wave (i.e., for the $i$th wave we have $w_{i}(x):=w(x,t=i\:T)$ where $T$ denotes the duration of the wave).
We update the weight between each wave using Eq.~(\ref{eqn:w_partial_meanfield}) and integrating over the time that the wave passes through each $x$-location:
\begin{eqnarray}
    \label{eqn:approx_weight_update_detailed}
    w_{i}(x) = w_{i-1}(x) + \tau_{\scriptscriptstyle +}\tau_{\scriptscriptstyle 3}A^{\scriptscriptstyle +}\hat{r}\int_{x/v}^{(x+2d)/v}r\left(r-\frac{\olsi{r}}{r_{\scriptscriptstyle 0}}\olsi{r}\right)dt.
\end{eqnarray}

Substituting in the expression for $\partial{w}/\partial{t}$ found in Eq.(\ref{eqn:w_partial_meanfield}) into Eq.(\ref{eqn:drdt_long}), we arrive at the equation for the change in firing rate of the V1 cell for the $i$th wave as follows
    \begin{eqnarray}
        \label{eqn:r_approx}
        \left\{\:
        \begin{aligned}
        \frac{\dd r}{\dd t} & = r(t)\left(r(t)-\frac{\olsi{r}(t)}{r_{\scriptscriptstyle 0}}\olsi{r}(t)\right) G\left(t\right) + F\left(t\right)\\
        G(t) & = g_0\:A^{\scriptscriptstyle +}\:\tau_{\scriptscriptstyle +}\tau_{\scriptscriptstyle \text{slow}}\hat{r}^{2}\int_{b(t)}^{u(t)}g(x)\dd x \\
	F(t) & \approx g_0\:\hat{r}\:g(x)\:w_i(x)\:\frac{\dd x}{\dd t}\:\Big|_{b(t)}^{u(t)} \\
        \end{aligned}\right.\kern-\nulldelimiterspace
    \end{eqnarray}

Together with Eq. (\ref{eqn:rbar_meanfield}) and Eq. (\ref{eqn:approx_weight_update_detailed}), we have arrived at the two-dimensional reduced model used in the text.
We note that the 3-d system does not qualitatively differ from the 2-d system by comparing the simulations' results.

\end{document}